\documentclass[%
headsepline=false,
abstract=true,
numbers=noendperiod
]{scrartcl}
\usepackage[UKenglish]{babel}                       
\usepackage[utf8]{inputenc}							
\usepackage[T1]{fontenc}							
\usepackage{lmodern}                                
\usepackage{fourier}								
\usepackage[section]{placeins}					    
\usepackage{booktabs}								
\usepackage{csquotes}                               
\usepackage{mathtools}								
\usepackage{amsmath}                                
\usepackage{amssymb}								
\usepackage{tensor}                                 
\usepackage{dsfont}                                 
\usepackage[dvipsnames]{xcolor}                     
\usepackage[mathscr]{euscript}					    
\usepackage{xfrac}                                  
\usepackage{tensor}                                 
\usepackage{stmaryrd}                               
\usepackage[%
backend=biber,
style=numeric,
citestyle=numeric-comp,
sorting=none,
doi=false,
isbn=false,
url=true,
maxbibnames=99
]{biblatex}                                         
\usepackage[pdfborder={0 0 0}]{hyperref}			
\usepackage{microtype}                              
\newcommand{\e}[1]{\operatorname{e}^{#1}}
\renewcommand{\i}{\operatorname{i}\!}
\renewcommand{\d}{\operatorname{d}\!}
\newcommand{\tr}{\operatorname{tr}}
\newcommand{\const}{\operatorname{const}}
\newcommand{\sl}{\mathfrak{sl}}

\newcommand{\iso}{\mathfrak{iso}}
\newcommand{\so}{\mathfrak{so}}
\newcommand{\reals}{\mathds{R}}
\newcommand{\scrim}{\mathpzc{m}}
\newcommand{\scris}{\mathpzc{s}}
\newcommand{\ad}{\operatorname{ad}}
\newcommand{\nslash}{\rotatebox[origin=c]{45}{$-$}}
\def\Xint#1{\mathchoice
{\XXint\displaystyle\textstyle{#1}}%
{\XXint\textstyle\scriptstyle{#1}}%
{\XXint\scriptstyle\scriptscriptstyle{#1}}%
{\XXint\scriptscriptstyle\scriptscriptstyle{#1}}%
\!\int}
\def\XXint#1#2#3{{\setbox0=\hbox{$#1{#2#3}{\int}$}
\vcenter{\hbox{$#2#3$}}\kern-.54\wd0}}
\def\fint{\Xint\nslash}
\DeclareMathAlphabet{\mathpzc}{OT1}{pzc}{m}{it}     
\numberwithin{equation}{section}
\KOMAoptions{parskip=never}
\begin{document}
\title{The Wilson Spool on Locally Flat Spacetimes\\ \ }
\author{Michel Pannier\textsuperscript{a,b}\thanks{\href{mailto:michel.pannier@unina.it}{michel.pannier@unina.it}}}
\publishers{%
\vspace{0.6cm}
\textsuperscript{\footnotesize a\ }\begin{minipage}[t]{0.8\linewidth}
    {\footnotesize \emph{Dipartimento di Fisica ``Ettore Pancini'', Università degli Studi di Napoli Federico II\\ Via Cintia, 21, I-80126 Napoli, Italy} \par}
\end{minipage}\\[0.1cm]
\textsuperscript{\footnotesize b\ }\begin{minipage}[t]{0.8\linewidth}
    {\footnotesize \emph{Istituto Nazionale di Fisica Nucleare (INFN), Sezione di Napoli\\ Via Cintia, 21, I-80126 Napoli, Italy} \par}
\end{minipage}\\[0.1cm]
}%
\date{}
\maketitle
\thispagestyle{empty}
\setcounter{page}{0}
\vspace{-0.4cm}
\begin{abstract}
This paper proposes a definition of what has previously been coined a \emph{Wilson Spool} in the case of three-dimensional gravity with vanishing cosmological constant. The de\-fi\-ni\-tion builds upon a construction of the one-loop partition function of a massive, spinning field from a fixed background holonomy. While the background is taken to be a flat-space cosmology, the definition of the Wilson spool is expected to hold independently of the underlying geometry. Some comments are given towards possible connections to its two-dimensional incarnations at non-vanishing cosmological constant.
\end{abstract}
\vspace{0.4cm}
\tableofcontents
\newpage
\section{Introduction}
In the challenge of understanding holography it has proven useful in the past to turn to lower-dimensional models, as they tend to give stronger control over both bulk and boundary theories. Indeed, in the case of three-dimensional general relativity in the bulk, the topological nature of the gravitational theory -- i.e. the absence of the graviton -- allows a consistent quantisation of the theory \cite{Achucarro:1986uwr,Witten:1988hc,Carlip:1998uc}. In many instances three-dimensional gravity is being used as an instructive toy model and its role may even go beyond that: it can be seen as the control group of quantum gravity in the sense that any features it shares with its higher-dimensional relatives are definitely not a consequence of the existence of a dynamical graviton field.

The motivation for the present work stems from the attempt to build holographic models in the case of vanishing cosmological constant, also known as flat-space holography \cite{Susskind:1998vk}. Here, the situation is somewhat less comfortable than in the case of AdS/CFT since the dual field theory, a Carrollian conformal field theory \cite{levy1965nouvelle,sen1966analogue,Barnich:2010eb,Bagchi:2010zz}, is comparably poorly understood and therefore the duality is still far from serving as a feasible definition of quantum gravity in asymptotically flat spacetimes. Accordingly, it is advisable to try and study both sides of the conjectured duality separately, at least as far as possible, and therefore this work deals exclusively with the (quantum-)gravitational side of the putative duality.

Although a full non-perturbative treatment of three-dimensional quantum gravity would have to include contributions of generic non-classical geometries, e.g. such featuring a degenerate metric, as well as a sum over topologies \cite{Witten:2007kt}, the approach taken here is a perturbative one. As such, the strategy is to extract quantum corrections around a classical saddle, eventually only considering a single contribution to the quantum-gravitational path integral. Starting point will be classical general relativity in three dimensions.

The full potential of three-dimensional classical Einstein gravity becomes apparent when it is formulated in the language of a Chern-Simons gauge theory,\footnote{Classically, the Einstein-Hilbert and Chern-Simons actions are equivalent modulo boundary terms. Accordingly, only gauge fields with a classical gravitational interpretation will be considered here.} in which the role of the underlying symmetries that govern the theory is particularly pronounced and its topological nature\footnote{There are no local degrees of freedom in pure three-dimensional Einstein gravity; still, there are non-trivial boundary degrees of freedom.} becomes rather transparent. However, this approach simultaneously raises a variety of new questions, for example concerning the (minimal) coupling of additional fields to the theory and how to formulate it in a purely algebraic language -- a question that is interesting both for practical purposes and from a conceptual point of view. Recently, an answer has been proposed in the case of massive, scalar and spinning fields for both positive and negative cosmological constant, introducing a gauge-invariant operator that has been coined the \emph{Wilson Spool} \cite{Castro:2023dxp,Castro:2023bvo,Bourne:2024ded,Fliss:2025sir,Bourne:2025azc}.

The idea is the following: In a gauge-theoretic language, objects that are capable of replacing the concept of geometry are Wilson lines, since they control parallel transport of minimally coupled fields, and it has been shown that they can in principle be thought of as the worldlines of massive particles \cite{Ammon:2013hba,Castro:2018srf}. Loosely speaking, wrapping such worldlines around a non-contractible cycle provides a technique for the calculation of the one-loop partition function \cite{Castro:2023dxp}, which motivates the approach to model the path integral by looping a Wilson line an arbitrary number of times and the result should be interpretable as the contribution of a quantum field, minimally coupled to the gauge theory. More precisely, the Wilson spool $\mathds{W}$ will calculate the partition function $Z_\text{M}$ of a massive matter field that lives on a fixed background geometry described in terms of a Chern-Simons gauge field $A$,
\begin{align}
    Z_\text{M}[A]=\e{\mathds{W}[A]}=\int\!\!\mathcal{D}\!\phi\,\e{\i S_\text{M}[\phi;A]}\,,
\end{align}
where $S_\text{M}[\phi;A]$ denotes the respective action of the minimally coupled matter field. In other words, the Wilson-spool operator is an object in which the path integral over the matter field has already been carried out and, as such, it presents itself as a suitable starting point to perform the path integral over the gauge sector and thus to capture quantum-gravitational effects. Interestingly, $\mathds{W}[A]$ depends merely on the background holonomy and the re\-pre\-sen\-ta\-tion that is associated with the field.

Immediately, one may wonder if this approach also applies in the case of vanishing cosmological constant and, given the group-theoretic nature of the Wilson spool as well as the fact that a Chern-Simons formulation is available for any value of the cosmological constant, the answer is expected to be affirmative. It is the purpose of the present paper to demonstrate that an analogous construction is possible and, indeed, the setup shares a number of characteristics with its negative-$\Lambda$ companion: the role of the (rotating) BTZ black hole as a fixed background possessing a finite temperature is played by so-called flat-space cosmologies and all information on the background geometry is stored in the holonomy. At the same time, the flat-space case does bring with it some peculiarities, such as the non-semisimple structure of the Poincaré group $\mathit{ISO}(2,1)$ and its algebra $\iso(2,1)$, which necessitates the use of induced representations. The latter are significantly less under control on a technical level than highest-weight representations, which are typically used in the case of the AdS$_3$-isometry algebra. In the present work, the explicit construction of an induced module is circumvented in so far as only a handful of necessary properties is assumed, just enough to efficiently calculate the (character of the) non-trivial background holonomy. However, apart from such variations stemming from the non-semisimple algebra structure, the construction of the flat-space Wilson spool and the one-loop partition function works along the same lines as in the curved cases and results in practically the same operator that has been presented in the previous literature \cite{Castro:2023dxp,Castro:2023bvo,Bourne:2024ded,Fliss:2025sir,Bourne:2025azc}, which is reassuring that the Wilson-spool proposal is fairly general.

Within section \ref{sec:foundations} of this paper I will briefly state the ingredients of the Chern-Simons gauge theory of three-dimensional gravity and discuss the appearance of Wilson-line operators as mediators of parallel transport of minimally coupled massive fields, both in the case of AdS and flat spacetime. That section furthermore contains relevant information on flat-space cosmologies, the induced representation module, and the (character of the) holonomy. The main result will be the construction of the Wilson-spool operator in section \ref{sec:Wilson_spool}, which builds upon the formulation of the one-loop partition function in terms of a weighted combination of traces of the holonomy, together with an appropriate renormalisation scheme. These calculations are performed for the case of flat-space cosmologies, albeit the general form of the Wilson spool is conjectured to be general. In the course of that construction, a known expression for the partition function of a massive, spinning field in the fixed background is reproduced. The same section provides a few comments on how a relation to previously proposed definitions of the Wilson spool in two-dimensional AdS and dS spaces could be established. As is common, the paper concludes with a discussion of its results and an outlook on possible future ventures in section \ref{sec:discussion}.
\section{The Chern-Simons Formulation and Matter Fields}\label{sec:foundations}
Starting point of the present considerations shall be a collection of some necessary ingredients to the minimal coupling of a matter field to three-dimensional gravity and its gauge-transformation behaviour. Discussing the coupling entirely in the language of the respective isometry algebras allows for a comparatively straightforward generalisation to any value of the cosmological constant.

In the next two subsections, the Wilson line for parallel transport of massive fields in the presence of the gravitational gauge fields is constructed, first in the case of negative, subsequently for vanishing cosmological constant. I will assume the reader to be familiar with the basic concepts of the analogy between the Einstein and Chern-Simons formulation \cite{Witten:1988hc,Carlip:1998uc}.
\subsection{Matter Fields at Negative Cosmological Constant}
It is instructive to start with a discussion of the case of negative cosmological constant. Most of this subsection will be a review, albeit with a particular focus on the transformation behaviour and parallel transport of massive fields, thereby paving the way towards a generalisation to the flat case and introducing a convenient notation.
\minisec{Isometry Algebra and Gauge Fields}
Three-dimensional Einstein gravity at negative cosmological constant may be described as a Chern-Simons gauge theory of the isometry algebra $\so(2,2)$, which in a relativistic basis is spanned by Lorentz generators $J_a$ and transvections $P_a$, where $a\in\{0,1,2\}$. The Lie algebra is defined by its Lie brackets
\begin{subequations}\label{eqs:Lie_bracket_all_Lambda}
\begin{align}
    \left[J_a\,,J_b\right]&=\varepsilon_{ab}{}^c J_c\,,\\
    \left[J_a\,,P_b\right]&=\varepsilon_{ab}{}^c P_c\,,\\
    \left[P_a\,,P_b\right]&=-\Lambda\varepsilon_{ab}{}^c J_c\,,
\end{align}
\end{subequations}
where the cosmological constant may be expressed in terms of the radius of curvature as $\Lambda=-1/l^2$ and in what follows it is $l=1$. Indices are moved using the metric in mostly-plus convention, i.e. $\eta=\operatorname{diag}(-1,1,1)$.

The Chern-Simons gauge field $A$ is an algebra-valued one-form that transforms under finite gauge transformations as\footnote{To be absolutely clear, here $\d g^{-1}\equiv \d(g^{-1})$.}
\begin{align}
    A\mapsto g A g^{-1}+g\d g^{-1}\,, && g=\e{\xi}\,, && \xi\in\so(2,2)\,,
\end{align}
and which can be decomposed as $A=\omega+e$ into spin connection $\omega=\omega^a J_a$ and vielbein $e=e^a P_a$. The gauge invariant equation of motion reads $\d A+A\wedge A=0$.

In many circumstances it may appear useful to employ the isomorphism\footnote{\label{fn:so_to_sl}%
One may first make the isomorphism $\so(2,2)\simeq\so(2,1)\oplus\so(2,1)$ manifest by diagonally embedding the $\so(2,1)$-generators $T_a$ like
\begin{align}
    J_a=\left(T_a\,,0\right)+\left(0\,,T_a\right)\,, && P_a=\left(T_a\,,0\right)-\left(0\,,T_a\right)\,.
\end{align}
Then the isomorphism $\so(2,1)\simeq\sl(2,\reals)$ can be realised explicitly as $L_{\pm 1}=T_0\mp T_1$ and $L_0=T_2$. Thus, we may think of any element $\xi\in\so(2,2)$ as being diagonally decomposed into $X,Y\in\sl(2,\reals)$, namely $\xi=(X\,,0)+(0\,,Y)$. As a short-hand, one may write $\xi=\xi^{(+)}+\xi^{(-)}$.
} %
$\so(2,2)\simeq\sl(2,\reals)\oplus\sl(2,\reals)$ that allows one to introduce two independent $\sl(2,\reals)$-valued gauge fields $A^{(\pm)}$, diagonally embedded like $A=(A^{(+)},0)+(0\,,A^{(-)})$. Each copy of the algebra is spanned by ge\-ne\-ra\-tors $L_m$ with $m\in\{0,\pm 1\}$ and defining Lie brackets
\begin{align}
    \left[L_m\,,L_n\right]=(m-n)L_{m+n}\,.
\end{align}
Naturally, the direct sum of Lie algebras becomes the direct product of Lie groups $\mathit{SO}(2,2)\simeq\mathit{SL}(2,\reals)\times\mathit{SL}(2,\reals)$ in the sense of their exponential subgroup, that is $g=(g_{(+)},g_{(-)})$, and the transformation behaviour of the individual gauge-field components reads
\begin{align}
    A^{(\pm)}\mapsto g_{(\pm)} A^{(\pm)} g_{(\pm)}^{-1}+g_{(\pm)}\d g_{(\pm)}^{-1}\,, && g_{(\pm)}=\e{\xi^{(\pm)}}\,, && \xi^{(\pm)}\in\sl(2,\reals)\,.
\end{align}
\minisec{Twist Operation and Matter Fields}
It is known, for example from the construction of worldline actions of massive particles \cite{Dzhordzhadze:1994np,Ammon:2013hba,Castro:2018srf} or the unfolding of massive fields \cite{Prokushkin:1998bq,Ammon:2011ua,Kessel:2018zqm}, that the respective matter fields transform inhomogeneously with respect to the two copies of $\sl(2,\reals)$ in the diagonal decomposition. Motivated by previous work on massive degrees of freedom in the context of higher-spin theory \cite{Vasiliev:1999ba,Alkalaev:2019xuv,Alkalaev:2020kut}, let us introduce a twist operation $\tau$, which is an involutive automorphism that is defined in terms of its action on the $\so(2,2)$-basis as $\tau(J_a)=J_a$ and $\tau(P_a)=-P_a$. Accordingly, acting on any diagonally decomposed Lie-algebra element it exchanges the two sectors,\footnote{In that sense, in the short-hand notation of footnote \ref{fn:so_to_sl} one may write $\tau\left(L^{(\pm)}_m\right)=L^{(\mp)}_m$.} $\tau(X,Y)=(Y,X)$.

Now, it shall be assumed that any massive (scalar or spinning) field $\phi(x)$ transforms under the action of the exponential subgroup $g=g(x)$ as
\begin{align}\label{eq:phi_transform_AdS}
    \phi\mapsto g\cdot\phi\cdot\tau\left(g^{-1}\right)\,.
\end{align}
The notation using the centre-dot product indicates that group elements act within some realisation of a unitary irreducible representation (UIR). When the matter field is diagonally decomposed as $\phi=(\phi^{(+)},\phi^{(-)})$, the individual field components transform like
\begin{align}
    \phi^{(\pm)}&\mapsto g_{(\pm)}\cdot\phi^{(\pm)}\cdot g_{(\mp)}^{-1}\,.
\end{align}
Apparently, in the presence of matter fields the two sectors of the gauge theory cannot be treated completely independently any more.
\minisec{Parallel Transport and the Wilson Line}
The possibility of gauge transformations makes it necessary to introduce parallel transport of matter fields, in order to make sense of objects such as the covariant derivative.\footnote{This is an ad-hoc argument; for a rigorous, axiomatic introduction in terms of fibre bundles see, e.g., \cite{Cherednikov:2020mtu}.} A parallel-transport operator of the field can generally be expressed using Wilson-line operators like so:
\begin{align}
    \operatorname{T}_{x_0\rightarrow x}\left[\phi(x_0)\right]=W^{\text{\tiny(left)}}(x,x_0)\cdot\phi(x_0)\cdot\tau\left(W^{\text{\tiny(right)}}(x_0,x)\right)\,,
\end{align}
which should transform under gauge transformations in the same way as $\phi(x)$. Thus, the transformation behaviour of the Wilson-line operators must be
\begin{align}\label{eq:W-AdS_transf_beh}
    W^{\text{\tiny(left/right)}}(x,x_0)\mapsto g(x)\cdot W^{\text{\tiny(left/right)}}(x,x_0)\cdot g(x_0)^{-1}\,,
\end{align}
which leads to identify both operators, $W^{\text{\tiny(left)}}(x,x_0)=W^{\text{\tiny(right)}}(x,x_0)\equiv W(x,x_0)$. Clearly, the Wilson-line operator is to be understood as an element of the gauge group and accordingly the respective Cartan-Maurer element has to be an element of the algebra,
\begin{align}\label{eq:W-AdS_diff_eq}
    \d W(x,x_0)W(x,x_0)^{-1}=B(x)\in\so(2,2)\,,
\end{align}
where the derivative acts on $x$. Then from \eqref{eq:W-AdS_transf_beh} follows the transformation behaviour of $B(x)$ and it is found to be that of $-A(x)$. Together with the initial-value condition $W(x_0,x_0)=\mathds{1}$ equation \eqref{eq:W-AdS_diff_eq} is (locally) the defining equation of the path-ordered exponential, and thus one arrives at the operator of parallel transport
\begin{align}\label{eq:transport_so}
    \operatorname{T}_{x_0\rightarrow x}\left[\phi(x_0)\right]=\left(\mathcal{P}\!\exp\int\limits_{x_0}^{x}\!(-A)\right)\,\cdot\,\phi(x_0)\,\cdot\,\tau\left(\mathcal{P}\!\exp\int\limits_{x}^{x_0}\!(-A)\right)\,.
\end{align}
More detailed information on definitions and conventions used here are collected in appendix \ref{app:sec:Wilson_lines}. The covariant derivative, which is defined in terms of parallel transport in \eqref{eq:def_cov_deriv}, then becomes
\begin{align}
    \operatorname{D}\!\phi(x)=\d\phi(x)+A(x)\cdot\phi(x)-\phi(x)\cdot\tau\left(A(x)\right)\,.
\end{align}

Alternatively, these expressions may be presented within the decomposition into $\sl(2,\reals)$-copies, in which the Wilson-line operator naturally decomposes into $W^{(\pm)}(x,x_0)$ in the same manner as the matter field such that each component transforms within a single copy of the group $\mathit{SL}(2,\reals)$, i.e. $W^{(\pm)}(x,x_0)\mapsto g^{(\pm)}(x)\cdot W^{(\pm)}(x,x_0)\cdot g^{(\pm)}(x_0)^{-1}$. Then the parallel-transport operator becomes $\operatorname{T}_{x_0\rightarrow x}[\phi^{(\pm)}(x_0)]=W^{(\pm)}(x,x_0)\cdot\phi^{(\pm)}(x_0)\cdot W^{(\mp)}(x_0,x)$ or, explicitly in terms of path-ordered exponentials,
\begin{align}\label{eq:transport_slsl}
    \operatorname{T}_{x_0\rightarrow x}\left[\phi^{(\pm)}(x_0)\right]=\left(\mathcal{P}\!\exp\int\limits_{x_0}^{x}\!\left(-A^{(\pm)}\right)\right)\,\cdot\,\phi^{(\pm)}(x_0)\,\cdot\,\left(\mathcal{P}\!\exp\int\limits_{x}^{x_0}\!\left(-A^{(\mp)}\right)\right)\,.
\end{align}
The covariant derivative takes on the well-known form
\begin{align}
    \operatorname{D}\!\phi^{(\pm)}(x)=\d\phi^{(\pm)}(x)+A^{(\pm)}(x)\cdot\phi^{(\pm)}(x)-\phi^{(\pm)}(x)\cdot A^{(\mp)}(x)\,.
\end{align}

Of both expressions, \eqref{eq:transport_so} and \eqref{eq:transport_slsl}, one may think of in an operational way: the outer right factor transports any object that is acted upon from $x$ to $x_0$ in a path-ordered manner, then $\phi(x_0)$ is applied and the result is transported back from $x_0$ to $x$, which eventually amounts to an action of $\phi(x)$. Clearly, the two paths taken between the two points are independent of each other.
\minisec{The Wilson Spool}
Now, being interested in parallel transport along a closed path around a non-contractible cycle one has to set $x=x_0$. In order to make direct contact with the holonomy around that cycle, the paths in the left and the right factor in \eqref{eq:transport_so} should be taken such that one of the paths shrinks to zero and only a single Wilson-line operator acts non-trivially; otherwise we would be going around the loop twice. Which of the paths becomes the loop and which one vanishes will not play any role.

Then single-valuedness of the field after transportation along the non-contractible (spatial) cycle can be taken as a starting point to construct the Wilson spool in AdS. This condition leads to restrictions on the poles of the partition function in terms of the respective representation weights, which -- after applying a suitable regularisation procedure -- result in the one-loop determinant of the massive (scalar) field and can eventually be generalised to an expression for the Wilson spool \cite{Castro:2023bvo,Castro:2023dxp,Bourne:2024ded},
\begin{align}\label{eq:Wilson_spool_AdS}
    \mathds{W}\left[A^{(\pm)}\right]=\frac{1}{4}\int\limits_{\mathscr{C}}\frac{\d\alpha}{\alpha}\coth\left(\frac{\alpha}{2}\right)\tr\left(\mathcal{P}\!\exp\left[\frac{\i\alpha}{2\pi}\oint\!\!A^{(+)}\right]\right)\tr\left(\mathcal{P}\!\exp\left[\frac{\i\alpha}{2\pi}\oint\!\!A^{(-)}\right]\right)\,.
\end{align}
Conventions on the normalisation, path ordering, as well as the integration variable have been adapted to the ones used throughout this paper and the contour $\mathscr{C}$ can be taken to consist of two different parts, one running from $-\infty$ to $+\infty$ slightly above the real $\alpha$-axis, the other one slightly below. The detailed construction will not be reviewed here.

As is obvious from the presence of the trace operation, the Wilson spool has to be thought of as being evaluated within some representation. Massive, spinning UIRs of $\so(2,2)$ are labelled by energy $\Delta$ and helicity $\scris$ and can be written as the tensor product of two lowest-weight representations of $\sl(2,\reals)$, the weights of which are fixed by $2h^{(\pm)}=\Delta\pm\scris$ \cite{Gunaydin:1986fe}. The expression \eqref{eq:Wilson_spool_AdS} strictly only holds for the scalar case $\scris=0$; for $\scris\ne 0$ one has to include an additional sum over the two possibilities to attach the $\sl(2,\reals)$-weights to the traces (the label ``$\pm$'' on the gauge fields is not to be confused with the one on the representation weights), but this is just a consequence of the splitting $\so(2,2)\simeq\sl(2,\reals)\oplus\sl(2,\reals)$ and could be avoided by working with the $(\Delta,\scris)$-representations of $\so(2,2)$, only.
\subsection{Matter Fields at Vanishing Cosmological Constant}\label{subsec:FS_prelims}
The formalisation of the transformation behaviour of minimally coupled matter fields presented in the previous subsection has the great advantage to be directly transferable to the case of vanishing cosmological constant. This subsection gives a comprehensive but self-contained overview of the formalism.
\minisec{Isometry Algebra and Gauge Fields}
Three-dimensional gravity in the case of asymptotically flat spacetimes can be put into the form of a Chern-Simons theory, as well. Here, the isometries are described in terms of the Poincaré algebra\footnote{Of course, relativistically $\iso(2,1)\simeq\so(2,1)\inplus\reals^3$ with Lie brackets as in \eqref{eqs:Lie_bracket_all_Lambda} for $\Lambda=0$, but here the isomorphism $\so(2,1)\simeq\sl(2,\reals)$ has been employed, see footnote \ref{fn:so_to_sl}; the abelian part is similarly written in an $\sl$-like basis.} $\iso(2,1)\simeq\sl(2,\reals)\inplus\reals^3$, spanned by Lorentz transformations $J_m$ and translations $P_m$, with $m\in\{0,\pm 1\}$ and commutation relations
\begin{subequations}
\begin{align}
    \left[J_m\,,J_n\right]&=(m-n)J_{m+n}\,,\\
    \left[J_m\,,P_n\right]&=(m-n)P_{m+n}\,,\\
    \left[P_m\,,P_n\right]&=0\,.
\end{align}
\end{subequations}
Due to the span of translations obviously forming a non-trivial ideal, this Lie algebra is not semi-simple.

The respective decomposition of the Chern-Simons gauge field is $A=\omega+e$ with spin connection $\omega=\omega^m J_m$ and vielbein $e=e^m P_m$; under a finite gauge transformation we again have $A\mapsto g A g^{-1}+g\d g^{-1}$. In principle, one may split any element of the algebra or its group into a Lorentz part and a translational part in order to account for the non-semi-simple nature of the theory. For example, group elements associated to finite gauge transformations can be decomposed as $g=g_{\text{L}}g_{\text{T}}$, where
\begin{align}\label{eqs:Lorentz_trans_split}
    g_{\text{L}}=\e{\xi_{\text{L}}}\,,\ \ \ g_{\text{T}}=\e{\xi_{\text{T}}} && \text{with} && \xi_{\text{L}}=\xi_{\text{L}}^m J_m\,,\ \ \ \xi_{\text{T}}=\xi_{\text{T}}^m P_m\,,
\end{align}
such that the individual gauge fields transform like
\begin{subequations}
\begin{align}
    \omega &\mapsto g_{\text{L}}\left(\omega+\d\,\right)g_{\text{L}}^{-1}\,,\\
    e &\mapsto g_{\text{L}}\left(e+\left[\xi_{\text{T}}\,,\omega\right]-\d\xi_{\text{T}}\right)g_{\text{L}}^{-1}\,.
\end{align}
\end{subequations}
Such a decomposition is not unique, of course, and other choices may be appropriate, depending on the circumstances.
\minisec{Twist Operation and Matter Fields}
Now, the transformation behaviour of a matter field will be assumed to contain a twist that switches the sign of translation generators, exactly as in the AdS case in\eqref{eq:phi_transform_AdS}, namely it will be of the form
\begin{align}
    \phi(x)\mapsto g\cdot\phi\cdot\tau\left(g^{-1}\right)\,,
\end{align}
where the centre-dot notation again indicates the action of group elements within some realisation of a representation that is not yet chosen. If written in a decomposed form, the twist operation simply acts like $\tau(g_{\text{L}}g_{\text{T}})=g_{\text{L}}g_{\text{T}}^{-1}$. Then a general ansatz to define parallel transport in terms of Wilson-line operators is
\begin{align}\label{eq:parallel_transport_operator}
    \operatorname{T}_{x_0\rightarrow x}\left[\phi(x_0)\right]=W^{\text{\tiny (left)}}(x,x_0)\cdot\phi(x_0)\cdot \tau\left(W^{\text{\tiny (right)}}(x_0,x)\right)\,.
\end{align}
Here, the action of $\tau(.)$ has been included on the right side for convenience but could alternatively be implicit in the definition of $W^{\text{\tiny (right)}}(x_0,x)$. In order for the parallel-transported field \eqref{eq:parallel_transport_operator} to transform the same way as $\phi(x)$ one has to demand
\begin{align}\label{eq:transformation_Wilson_line}
    W^{\text{\tiny (left/right)}}(x,x_0)&\mapsto g(x)\cdot W^{\text{\tiny (left/right)}}(x,x_0)\cdot g(x_0)^{-1}\,,
\end{align}
such that we may identify left and right Wilson-line operators, $W^{\text{\tiny (left)}}(x,x_0)=W^{\text{\tiny (right)}}(x,x_0)\equiv W(x,x_0)$. From here on the steps summarised in appendix \ref{app:sec:Wilson_lines} can be applied: demanding the Wilson-line operator to produce a Cartan-Maurer form that takes values in the Lie algebra, imposing the initial-value condition $W(x_0,x_0)=\mathds{1}$ (on the class of paths that do not contain closed loops), and using again the transformation behaviour \eqref{eq:transformation_Wilson_line}, leads to the explicit expression in terms of the path-ordered exponential of the gauge fields
\begin{align}
    W(x,x_0)=\mathcal{P}\!\exp\int\limits_{x_0}^x\! (-\omega-e)=\overline{\mathcal{P}}\!\exp\int\limits_{x}^{x_0}\! (\omega+e)\,.
\end{align}
The alternative form on the right-hand side is written in terms of an anti-path-ordered exponential -- conventions for path ordering and anti-path ordering are stated in equations \eqref{eqs:path_ordering_conv}.

Taking everything together, an explicit way of writing the parallel-transport operator acting on a massive, spinning field reads
\begin{subequations}
\begin{align}
    \operatorname{T}_{x_0\rightarrow x}\left[\phi(x_0)\right]&=\left(\mathcal{P}\!\exp\int\limits_{x_0}^{x}\!(-\omega-e)\right)\,\cdot\,\phi(x_0)\,\cdot\,\left(\mathcal{P}\!\exp\int\limits_{x}^{x_0}\!(-\omega+e)\right)\label{eq:parallel_transport_FSa}\,,\\
    &=\left(\overline{\mathcal{P}}\!\exp\int\limits_{x}^{x_0}\!(\omega+e)\right)\,\cdot\,\phi(x_0)\,\cdot\,\left(\overline{\mathcal{P}}\!\exp\int\limits_{x_0}^{x}\!(\omega-e)\right)\label{eq:parallel_transport_FSb}\,.
\end{align}
\end{subequations}
Let me collect a few comments on this expression in the following.
\begin{itemize}
    \item A particular realisation of a massive, spinning induced representation of the Poincaré group is yet to be chosen, as is still indicated by the centre-dot product.
    \item The paths in the left and the right factor are independent of each other and only need to have the same endpoints. The path-ordered exponentials do not depend on the precise shape of these paths but only on their respective homotopy class (i.e. the Wilson line gives the same result for paths of fixed endpoints that can be continuously deformed into each other).
    \item The Wilson-line operator can be decomposed into a Lorentz and a translational part like $W(x,x_0)=W_\text{L}(x,x_0)W_\text{T}(x,x_0)$, such that it becomes apparent that path ordering is associated with the non-abelian part, only. In that case
    \begin{align}
        W_\text{L}(x,x_0)=\mathcal{P}\!\exp\int\limits_{x_0}^x\! (-\omega)\,, && W_\text{T}(x,x_0)=\e{-\int\limits_{x_0}^x W_\text{L}(x_0,x')\cdot e(x')\cdot W_\text{L}(x',x_0)}\,.
    \end{align}
    \item The covariant derivative is defined in terms of parallel transport as before, see equation \eqref{eq:def_cov_deriv}, and takes the expected form\footnote{Note that the appearance of the adjoint or twisted-adjoint \emph{action} does not directly imply an appearance of the adjoint or twisted-adjoint \emph{representation}. The matter field is not considered to be an element of the Lie algebra (or a quotient of its universal enveloping algebra) but rather of the representation space of the respective UIR, on which said action is carried out. That representation space may be thought of as some infinite-dimensional function space, as sketched in the minisection on induced representations below.}
    \begin{subequations}
    \begin{align}
        \operatorname{D}\!\phi(x)&=\d\phi(x)+\!\tensor*[^{\tau}]{\ad}{_{A(x)}}\left(\phi(x)\right)\\
        &=\d \phi(x)+\left[\omega(x)\,,\phi(x)\right]+\left\{e(x)\,,\phi(x)\right\}\,,
    \end{align}
    \end{subequations}
    where the definition of the twisted-adjoint action \cite{Alkalaev:2019xuv,Alkalaev:2020kut} is $\tensor*[^{\tau}]{\ad}{_X}(Y):=X\cdot Y-Y\cdot \tau(X)$. Note that $\operatorname{D}\!\phi(x)$ transforms in an adjoint way under simultaneous gauge transformations of $A(x)$ and $\phi(x)$, as it should.
\end{itemize}
\minisec{Flat-Space Cosmologies in Stabiliser Gauge}
In (outgoing) Eddington-Finkelstein coordinates $(u,r,\varphi)$ a common definition of asymptotic flatness (i.e. fixing a gauge and fall-off conditions) leads to a class of general solutions to Einstein's equations that can be written in terms of the gauge fields as \cite{Cornalba:2002fi,Barnich:2012aw,Afshar:2013bla}
\begin{subequations}\label{eqs:spin-connection_vielbein}
\begin{align}
    \omega&=\left(J_1-\frac{M(\varphi)}{4}J_{-1}\right)d\!\varphi\,,\\
    e&=\left(P_1-\frac{M(\varphi)}{4}P_{-1}\right)d\!u+\frac{1}{2}P_{-1}d\!r+\left(r P_0-\frac{N(u,\varphi)}{2}P_{-1}\right)d\!\varphi\,.
\end{align}
\end{subequations}
The functions $M(\varphi)$ and $N(u,\varphi)$ are connected through the integrability condition $\partial_\varphi M(\varphi)=2\partial_u N(u,\varphi)$ but unconstrained otherwise. As a convenient first step, the radial dependence may be entirely eliminated by a large gauge transformation, namely $A=g_r^{-1}A^{(r=0)}g_r-\d g_r^{-1}g_r$ with $g_r=\exp(\sfrac{r}{2}\,P_{-1})$.

Since the Wilson-spool calculation is supposed to be a perturbative one around a classical saddle, only the case of so-called flat-space cosmologies is of importance here; for these the charges are constants, namely $M(\varphi)=M>0$ and $N(u,\varphi)=N\ne 0$. For that case, we are interested in the $\varphi$-components\footnote{Generally, in the Wilson loop around the non-contractible cycle we can use any circle that encloses the singularity; but of course the choice $u=\const$, $r=\const$ is particularly convenient because the path-ordering becomes obsolete and we are able to perform the angular integral.} of the two possible combinations of the gauge fields \eqref{eqs:spin-connection_vielbein}, after having removed the $r$-dependency, which read
\begin{align}
    \omega_\varphi^{(r=0)}\pm e_\varphi^{(r=0)}=J_1-\frac{M}{4}J_{-1}\mp\frac{N}{2}P_{-1}\,.
\end{align}
As it turns out, this can be generated by a gauge transformation from the Lie-subalgebra $\mathfrak{s}\subset\iso(2,1)$ spanned by $J_0$ and $P_0$,
\begin{align}\label{eq:stabiliser_gauge}
    g^{-1}\left(\sigma J_0\pm\lambda P_0\right)g=J_1-\frac{M}{4}J_{-1}\mp\frac{N}{2}P_{-1}\,,
\end{align}
with coefficients $\sigma$ and $\lambda$ yet to be determined. The abelian Lie algebra $\mathfrak{s}$ generates the Lie-subgroup $S\subset\mathit{ISO}(2,1)$ and it makes sense to perform the respective coset decomposition, $g^{-1}=\Phi s$, where $s\in S$. Further, decomposing $\Phi=g_{\text{L}}g_{\text{T}}$ in the sense of \eqref{eqs:Lorentz_trans_split} we arrive at the equations
\begin{subequations}
\begin{align}
    \sigma g_{\text{L}}\,J_0 g_{\text{L}}^{-1}&=J_1-\frac{M}{4}J_{-1}\,,\\
    g_{\text{L}}\left(\pm\lambda P_0+\sigma\left[\xi_{\text{T}}\,,J_0\right]\right)g_{\text{L}}^{-1}&=\mp\frac{N}{2}P_{-1}\,.
\end{align}
\end{subequations}

A quick way to obtain a solution for $\sigma$ and $\lambda$ is by application of a convenient bilinear form\footnote{It appears useful to introduce a bilinear form on the Poincaré algebra that generalises the one introduced in \cite{Witten:1988hc} and is defined through \cite{Pannier:2023srn}
\begin{subequations}\label{eqs:blf}
\begin{align}
    \left\langle J_m\,,J_n\right\rangle&=0\,,\\
    \left\langle J_m\,,P_n\right\rangle&=\scrim\scris\,\eta_{m\!n}\,,\\
    \left\langle P_m\,,P_n\right\rangle&=\scrim^2\eta_{m\!n}\,,
\end{align}
\end{subequations}
where $(\eta_{m\!n})=\operatorname{antidiag}(-2,1,-2)$ and the constants $\scrim$ and $\scris$ are included to amount for the length dimension contained in the translation generators but are of no further significance here. (The notation is supposed to remind of the mass squared and spin Casimir elements $\mathscr{M}^2$ and $\mathscr{S}$, to which this form is related -- these may be parametrised as $\mathscr{M}^2\simeq\scrim^2\mathds{1}$ and $\mathscr{S}\simeq \scrim\scris\mathds{1}$.) This form is invariant under the adjoint action of Lorentz transformations but not translations.} to the equations and by noting that the objects $g_{\text{L}}\,J_0 g_{\text{L}}^{-1}$ and $g_{\text{L}}\,P_0 g_{\text{L}}^{-1}$ must have the same expansion into basis elements upon replacing $J_m\mapsto P_m$. This results (for both signs) in
\begin{align}
    \sigma^2=M && \text{as well as} && \sigma\lambda=N\,.
\end{align}
The ambiguity in the sign of $\sigma$ may be formulated in a form that relates it to the sign of the function $N$ by writing $\operatorname{sgn}(\sigma)=\pm\operatorname{sgn}(N)$, such that $\operatorname{sgn}(\lambda)=\pm 1$.\footnote{\label{fn:comparison_sigma_lambda}Comparison to the metric in Schwarzschild-like coordinates shows that one may identify $|\lambda|=r_0$, the latter being the cosmological horizon; then in the notation of \cite{Bagchi:2013lma,Bagchi:2013qva}, $\sigma=\pm\hat{r}_+$, which has an interpretation as a remnant of the re-scaled outer horizon of a BTZ black hole that is being pushed to infinity when taking the limit of vanishing cosmological constant.} The explicit form of $g_{\text{L}}$ and $g_{\text{T}}$ will not be needed but stated for completeness; using a matrix representation one may easily find
\begin{align}
    g_{\text{L}}=\exp\left(\frac{\sigma}{2}J_{-1}\right)\exp\left(\frac{1}{\sigma}J_1\right)\,, && g_{\text{T}}=\exp\left(\pm\frac{\lambda}{2}P_{-1}\right)\exp\left(\mp\frac{\lambda}{2\sigma^2}P_1\right)\,,
\end{align}
which is of course not unique but a representative of an equivalence class of solutions.
\minisec{Induced Representation}
Up to now it has not been specified in which unitary irreducible representation (UIR) of the Poincaré group or algebra the action of, e.g., the Wilson-line operator takes place, only that we are dealing with a massive, spinning UIR. Recall that unitary representations of non-semisimple groups, such as Poincaré, are necessarily induced rather than highest- or lowest-weight representations \cite{Campoleoni:2016vsh,Oblak:2016eij}.

Without fully constructing representation space here, I will assume to be in possession of a UIR of $\iso(2,1)$ labelled by mass $\scrim>0$ and spin (helicity) $\scris\in\reals$. The parameter $\scris$ is not necessarily quantised in three spacetime dimensions, such that $\phi(x)$ may be regarded as an anyonic -- and in that sense \emph{spinning} -- field. The action of generators within the representation will be indicated by small letters $j_m$, $p_m$ and a concrete realisation will be thought of as living on the coadjoint orbit (aka. phase space) leading to a multiplicative action of translations and a well-defined inner product (see, e.g., the construction in \cite[section 6.2]{Pannier:2023srn}).

To construct the induced module, start with a $j_0$-eigenfunction and span a space of functions generated from repeated application of boosts:
\begin{align}
    j_0\cdot f_0=\scris\,f_0\,, && \tilde{f}_{mn}\sim(j_1)^m(j_{-1})^n\cdot f_0\,.
\end{align}
The reference function $f_0$ has to be chosen in such a way that convergence of the integral that defines an inner product (given in terms of the invariant measure on the coadjoint orbit) is ensured. Then one may construct an orthonormal basis from the standard Gram-Schmidt procedure, leading to linear combinations of the $\tilde{f}_{mn}$ with fixed $m-n$. That is, there exists a family of special functions $f_{mn}$ that are orthonormal with respect to an inner product,
\begin{align}
    \left(f_{mn}\,,f_{m'n'}\right)=\delta_{mm'}\delta_{nn'}\,,
\end{align}
and fulfil an eigenvalue equation that follows directly from the commutation rules of $j_0$ with powers of the Lorentz boosts $j_{\pm 1}$,
\begin{align}
    j_0\cdot f_{mn}=(\scris-m+n)f_{mn}\,.
\end{align}
The $j_{\pm 1}$ themselves act as differential operators that shift the indices of basis functions, while translations will -- as stated above -- act multiplicatively in such a representation. If restricted to the rest frame, the action of $p_{\pm 1}$ is necessarily vanishing while $p_0=\scrim \mathds{1}$.

Note that the construction automatically parametrises the second-order Casimir elements as multiples of the identity, meaning that the equations
\begin{subequations}\label{eqs:Casimirs}
\begin{align}
    \mathscr{M}^2&\equiv p_0 p_0-p_1 p_{-1}=\scrim^2\mathds{1}\,,\\
    \mathscr{S}&\equiv j_0 p_0-\frac{1}{2}\left(j_1 p_{-1}+j_{-1}p_1\right)=\scrim\scris\mathds{1}
\end{align}
\end{subequations}
are identically fulfilled for these operators.

Finally, from this basis one is able to define a trace operation in the usual way, namely
\begin{align}\label{eq:trace}
    \tr(X):=\sum_{m,n=0}^\infty \left(f_{mn}\,,X\cdot f_{mn}\right)\,.
\end{align}
\minisec{Holonomy and Character}
Flat-space cosmologies are geometries that contain a non-contractible cycle and, accordingly, non-trivial holonomy around that cycle. The holonomy operator may, loosely speaking, be obtained by taking a Wilson-line operator and closing the path of integration around such a non-contractible cycle, meaning
\begin{align}
    \operatorname{Hol}_\mathcal{C}[\omega,e]=\mathcal{P}\!\exp\oint\limits_\mathcal{C}(-\omega-e)=\overline{\mathcal{P}}\!\exp\oint\limits_\mathcal{C}(\omega+e)\,.
\end{align}
Here, the closed curve $\mathcal{C}$ can be parametrised by the angular coordinate $\varphi$ while keeping $u$ and $r$ fixed. Since the gauge fields are constant, (anti-)path ordering becomes obsolete and the integration can be carried out; the distinction between path ordering and anti-path ordering simply results in a sign ambiguity that results from the two possible directions in which the $\varphi$-circle can be parametrised. To avoid clutter, let me from here on choose a positive sign in the exponent.

The transformation behaviour of the holonomy is known by its construction as a closed Wilson line and we can pass to the particular gauge discussed above, resulting in
\begin{align}\label{eq:holonomy}
    \operatorname{Hol}[\omega,e]=g_r^{-1}\e{2\pi\left(\omega_\varphi^{(r=0)}+e_\varphi^{(r=0)}\right)}g_r=g_r^{-1} g^{-1} \e{2\pi\left(\sigma J_0+\lambda P_0\right)}g g_r\,.
\end{align}
Note that none of the group elements depends on $\varphi$, such that here there are no issues in relation to path ordering. If we were to include the case of Minkowski spacetime (which at this point necessitates complexification), $\sigma=\pm\i$ and $\lambda=0$, then the holonomy indeed becomes the unit element.

Using the definition of a trace in the massive, spinning representation introduced above it is possible to obtain the character of the holonomy, reading
\begin{align}\label{eq:ch_holonomy}
    \tr\left(\operatorname{Hol}[\omega,e]\right)=-\frac{\e{2\pi(\scris\sigma+\scrim\lambda)}}{4\sinh^2(\pi\sigma)}\,,
\end{align}
which is of the expected form of Poincaré characters. In the case of vacuum Minkowski spacetime the trace becomes infinite.
\section{The Wilson-Spool Proposal}\label{sec:Wilson_spool}
Within this section I will perform the construction of the Wilson spool for the case of asymptotically flat spacetimes, roughly following the ideas that were laid out in \cite{Castro:2023bvo,Castro:2023dxp}. First, in subsection \ref{subsec:partition_fct} the one-loop partition function will be re-constructed from its poles and identified with a weighted integral over the trace of the holonomy, thereby first regulating the UV divergence and in a second step the sum over quantum numbers. The proposed definition of the Wilson spool is given in \ref{subsec:spool_def}, before concluding with a couple of remarks on the Wilson spool in different dimensions.
\subsection{The Partition Function from Holonomies}\label{subsec:partition_fct}
Starting point of the construction is the analyticity of the partition function that allows one to construct the one-loop contribution from the knowledge of its poles \cite{Denef:2009kn}. From the viewpoint of representation theory this allows us to simply find all conditions on the quantum numbers involved in a solution of the field equations and assemble the function that has precisely these poles.

Recall that the Casimir equations are supposed to be trivially fulfilled within the representation considered here, as mentioned around equations \eqref{eqs:Casimirs}. Furthermore, global regularity of the physical field content is ensured if the representation of the Lie algebra $\sl(2,\reals)\inplus\reals^3$ exponentiates to a representation of (the universal cover of) the respective group, which is the case for the $\sl(2,\reals)$-subalgebra \cite{Kitaev:2017hnr} and trivially so for the abelian part.
\minisec{From Single-Valuedness to Poles}
Single-valuedness of the solution will be a necessary condition, meaning that the field is not allowed to change its value under transport around the non-contractible cycle,
\begin{align}
    \left(\mathcal{P}\!\exp\oint\limits_{\mathcal{C}^+}\!(-\omega-e)\right)\,\cdot\,\phi(x)\,\cdot\,\left(\mathcal{P}\!\exp\oint\limits_{\mathcal{C}^-}\!(-\omega+e)\right)\stackrel{!}{=}\phi(x)\,.
\end{align}
When closing the contour, one of the two paths $\mathcal{C}^\pm$ should shrink to zero, otherwise we would wind around the same cycle twice; in the following choose $\mathcal{C}^+$ as the circle parametrised by $\varphi$ and take $\mathcal{C}^-$ to be trivial -- the other option may in principle be recovered by application of the twist operation $\tau(.)$ but the following considerations will not depend on that choice.\footnote{As soon as a single Wilson-line operator is isolated the respective sign choice is inconsequential since the inner automorphism $P_m\mapsto -P_m$ can be regarded as a re-scaling of generators.} Knowing how the parallel-transport operator behaves under gauge transformations, we find this condition to be invariant under the gauge transformation given in \eqref{eq:stabiliser_gauge}, such that we have\footnote{Furthermore, here the orientation of the cycle has been chosen as to compensate for the minus sign in the exponential.}
\begin{align}
    \e{2\pi\left(\sigma J_0+\lambda P_0\right)}\cdot\,\phi(x)\stackrel{!}{=}\phi(x)\,.
\end{align}
For the calculations that are following it will moreover be necessary to work with the Euclideanised spacetime. The known Wick rotation on the level of the metric \cite{Bagchi:2013lma} not only replaces $t\equiv \i t_{\text{E}}$ but also $M\equiv -M_{\text{E}}$ and $N\equiv -\i N_{\text{E}}$. For the expression at hand we have to set $\sigma=-\i\sigma_{\text{E}}$ and $P_0=-\i P_0^{\text{E}}$ (this generator being related to time translations), while $\lambda$ remains unaltered.\footnote{As noted in footnote \ref{fn:comparison_sigma_lambda}, $\lambda$ may be identified with the locus $r_0$ of the cosmological horizon. As for the relation of $P_0$ in the $\sl$-like basis with time translation, it is to mention that the isomorphism presented in footnote \ref{fn:so_to_sl} does not make this transparent; one needs to consider the complexification of algebras to understand the correct physical interpretation of generators.} All in all, the Euclidean single-valuedness condition reads
\begin{align}\label{eq:single_value_eucl}
    \e{-2\pi\i\,\left(\sigma_{\text{E}} J_0+ \lambda P_0^{\text{E}}\right)}\cdot\,\phi(x)\stackrel{!}{=}\phi(x)\,.
\end{align}

Now, the induced representation discussed in subsection \ref{subsec:FS_prelims} shall be used. On the orthonormal basis introduced there the action of $J_0$ becomes an eigenvalue equation and the generator $P_0$ will act multiplicatively; for notational simplicity I will assume the rest-frame condition, in which the eigenvalue of the latter is the mass $\scrim$ -- there is however no loss of generality. Then the condition on the exponential in equation \eqref{eq:single_value_eucl} becomes
\begin{align}
    \sigma_{\text{E}}\left(\scris-m+n\right)+\lambda\scrim_{\,\text{E}}\stackrel{!}{=}k\,, && k\in\mathds{Z}\,.
\end{align}
This equation dictates the simple poles of the partition function so that we may write it as a product over all possible values of $m$, $n$, and $k$. However, both possible sign choices in $\sigma=\pm\operatorname{sgn}(N)\sqrt{M}$ (which implies $\lambda=\pm|\lambda|$) have to be taken into account; from now on fixing the notation such that $\sigma$ indicates the identification with the upper sign leads to
\begin{align}
    Z^2=\prod_{k\in\mathds{Z}}\prod_{m,n\in\mathds{N}_0}\big(|k|+\sigma_{\text{E}}(\scris-m+n)+|\lambda|\scrim_{\,\text{E}}\big)^{-1}\big(|k|-\sigma_{\text{E}}(\scris-m+n)-|\lambda|\scrim_{\,\text{E}}\big)^{-1}\,.
\end{align}
\minisec{UV-Regularisation}
Now we may return to Lorentzian signature, $\sigma_{\text{E}}=\i\sigma$ and $\scrim_{\,\text{E}}=\i\scrim$. Taking the logarithm turns all products into sums and, using the integral representation
\begin{align}
    \ln(x)=\int\limits_0^\infty\frac{\d\alpha}{\alpha}\left(\e{-\alpha}-\e{-\alpha x}\right)\,, && x\in\mathds{C}\setminus(-\infty,0]\,,
\end{align}
the expression becomes
\begin{align}
    \ln(Z)=-\sum_{m,n\in\mathds{N}_0}\ \sum_{k\in\mathds{Z}}\ \int\limits_0^{\infty}\frac{\d\alpha}{\alpha}\left(\e{-\alpha}-\e{-\alpha |k|}\cos\big(\alpha\sigma(\scris-m+n)+\alpha|\lambda|\scrim\big)\right)\,.
\end{align}
The sum over $k$ must be regularised, apparently. That can be achieved by including a damping factor $\exp(-\varepsilon |k|)$ and, after performing the sum, subtraction of the divergent piece in the limit $\varepsilon\rightarrow 0$. However, one is forced to set the lower integration limit to $\varepsilon$ at the same time, in order to avoid introducing an additional divergence. Then:
\begin{subequations}
\begin{align}
    \ln(Z)&\simeq \sum_{m,n\in\mathds{N}_0}\lim_{\varepsilon\searrow 0}\,\int\limits_{\varepsilon}^\infty \frac{\d\alpha}{\alpha}\coth\left(\frac{\alpha}{2}\right)\cos\big(\alpha\sigma(\scris-m+n)+\alpha|\lambda|\scrim\big)\,.
\end{align}
\end{subequations}
Symmetrisation of the integration range eventually leads to a principal-value integral; furthermore, the sum over $m$ and $n$ may be identified as the trace operation \eqref{eq:trace} and the final expression becomes
\begin{align}\label{eq:lnZ}
    \ln(Z)\simeq\frac{1}{2}\ \fint\limits_{\mathds{R}}\frac{\d\alpha}{\alpha}\coth\left(\frac{\alpha}{2}\right)\tr\left(\e{\i\alpha\left(\sigma J_0+|\lambda| P_0\right)}\right)\,.
\end{align}
Note that after replacing $\scrim$ with $P_0$ we have essentially undone the rest-frame assumption, which is not surprising since the multiplicative action of the generator could as well have been carried through the calculation. At this point it is already transparent from comparison to equation \eqref{eq:holonomy} that the partition function is essentially obtained from the trace of the respective holonomy, i.e. the character within a specific representation.

Regularisation of the $k$-sum is expected to correspond to a regularisation of the UV divergence, while the divergent trace is still present. A priori it is not clear whether or not these two infinite sums may be interchanged and which one is to be regularised first; for the present work it shall be sufficient that a correct expression for the partition function is successfully reproduced later.
\minisec{Defining the Principal-Value Integral}
It is not immediately obvious that the principal-value integral written in \eqref{eq:lnZ} is actually well defined. Let us pull the integral inside of the trace and abbreviate the Lie-algebra element as $X\equiv \sigma J_0+|\lambda| P_0$; then the integral under question is
\begin{align}
    \fint\limits_{\mathds{R}}\!\d\alpha\,f(\alpha)=\lim_{\varepsilon\searrow 0}\left(\int\limits_{-\infty}^{-\varepsilon}\d\alpha+\int\limits_{\varepsilon}^{\infty}\d\alpha\right)f(\alpha)\equiv\lim_{\varepsilon\searrow 0}\left(I_\varepsilon\right)\,, && f(\alpha)=\frac{\coth\left(\frac{\alpha}{2}\right)}{\alpha}\e{\i\alpha X}\,.
\end{align}
To evaluate $I_\varepsilon$, the points $-\varepsilon$ and $\varepsilon$ can be connected by a semi-circle in the complex $\alpha$-plane that bypasses the origin within the upper or lower complex half plane -- let me first discuss the second option for concreteness, i.e. negative imaginary part of $\alpha$ -- and the integration path may be closed by connecting $\infty$ and $-\infty$ by a semi-circle of infinite radius. If the latter one is taken on the upper or lower complex half plane will depend on the behaviour of the exponential at infinite radius. Then the value of the integral along the closed contour is given by the sum of the residues of the enclosed poles, which are positioned along the imaginary axis at $\alpha_0=0$ as well as $\alpha_k=2\pi\i k$, where $k\in\mathds{Z}\setminus\{0\}$. Concretely,
\begin{align}
    I_\varepsilon+I_{\text{circ}}+I_\infty=2\pi\i\sum_{k=0}^\infty \operatorname{Res}_k(f)
\end{align}
and the values of these residues are
\begin{align}
    \operatorname{Res}_0(f)=2\i X\,, && \operatorname{Res}_k(f)=\frac{1}{\i\pi}\frac{\e{-2\pi k X}}{k}\,.
\end{align}
The exponential at this point is a formal expression in terms of the Lie-algebra generators but we will have to assume its possessing a fall-off as $|\alpha|\rightarrow\infty$ in either the upper or lower complex plane, since both measure and weight stay finite in that limit; only then, $I_\infty=0$. The integral around the semi-circle can be calculated directly by parametrising $\alpha=\varepsilon\exp(\i\varphi)$ and expanding both the \emph{cotangens hyperbolicus} and the exponential into a power series in the phase factor, which results in $I_{\text{circ}}=-2\pi X$. Combining these results leads to
\begin{align}\label{eq:Iepsilon}
    I_\varepsilon=2\sum_{k=1}^\infty\frac{\e{-2\pi k X}}{k}-2\pi X\,.
\end{align}
Apparently, the limit $\varepsilon\rightarrow 0$ becomes trivial. Furthermore, it is straightforward to show that the choice of the $\varepsilon$-circle lying in the upper half plane (and the residue sum thus starting at $k=1$) yields exactly the same result and we have thus demonstrated that the $\alpha$-integral is well defined, i.e. the renormalisation has successfully removed the divergence stemming from the sum over $k$ above.
\minisec{Regularisation of the Trace -- Fixing the Integration Contour}
Now let us focus on the divergence that originates from the trace operation. When the Lie-algebra element $X$ is replaced with its eigenvalue in the induced representation and the sum over quantum numbers is inserted, the explicit appearance of $X$ outside of an exponential in \eqref{eq:Iepsilon} apparently introduces a diverging sum. Thus, a renormalisation of the trace necessarily forces the integration contour to exclude the singularity at $\alpha=0$. But there are further singularities that can only be seen after explicit evaluation of the trace, which gives
\begin{align}\label{eq:trace_alpha}
    \tr\left(\e{\i\alpha(\sigma J_0+\lambda P_0)}\right)=\frac{\e{\i\alpha(\sigma\scris+\lambda\scrim)}}{4\sin^2(\alpha \sigma)}\,,
\end{align}
leading to a family of poles sitting along the real axis. One may thus fix an integration contour that runs parallel to the real $\alpha$-axis from $-\infty$ to $\infty$ but slightly shifted towards positive or negative imaginary values of $\alpha$. Which option to choose is determined by demanding fall-off of \eqref{eq:trace_alpha} together with the weight function at $|\alpha|\rightarrow\infty$; careful analysis of the expression leads to the condition
\begin{align}
    1+\kappa\operatorname{sgn}\left[\operatorname{Im}(\alpha)\right]>0\,, && \kappa\equiv\frac{\sigma\scris+|\lambda|\scrim}{2|\sigma|}\,.
\end{align}
This implies that for $\kappa\ge 1$ the upper half plane has to be chosen, while for $\kappa\le -1$ the lower half plane is enforced. Accordingly, there is a region of parameters $|\kappa|<1$, where there is no immediate restriction and exponential fall-off is ensured in both half planes; it can however be seen that the two choices do not lead to the same result in the integration later, but we do want the parameter $\kappa$ to be continuously tunable, in principle. The resolution of that is fixing the integration curve by taking the sign of the imaginary part of $\alpha$ to be the sign of $\kappa$. Accordingly, denote by $\mathscr{C}_\pm$ the integration contour slightly above/below the real axis.

The expression for the partition function reads (from now on dropping the indication of regularisation as `$\simeq$')
\begin{align}\label{eq:part_fct_reg_complete}
    \ln(Z)=\frac{1}{8}\int\limits_{\mathscr{C}_\pm}\!\frac{\d\alpha}{\alpha}\coth\left(\frac{\alpha}{2}\right)\frac{\e{\i\alpha\left(\sigma\scris+|\lambda|\scrim\right)}}{\sin^2\left(\alpha\sigma\right)}\equiv\frac{1}{8}\int\limits_{\mathscr{C}_\pm}\!\!\d\alpha\,f(\alpha)
\end{align}
and its poles along the imaginary axis are lying at $\alpha_k=2\pi\i k$, $k\in\mathds{Z}$. The respective residues are given as
\begin{align}
    \operatorname{Res}_{k}(f)=\frac{\i}{\pi}\frac{\e{-2\pi k\left(\sigma\scris+|\lambda|\scrim\right)}}{k\sinh^2\left(2\pi\sigma k\right)}\,, && k\in\mathds{Z}\setminus\{0\}
\end{align}
and, as just discussed, the integral can be obtained in terms of the sum of residues of $k>0$ or $k<0$, depending on the sign of $\sigma\scris+|\lambda|\scrim$. All in all, the result for the one-loop partition function reads
\begin{align}
    \ln(Z)=\mp\frac{1}{4}\sum_{k=1}^\infty \frac{\e{-2\pi k \big|\sigma\scris+|\lambda|\scrim\big|}}{k\sinh^2(2\pi\sigma k)}\,,
\end{align}
where the sign in front is to be understood as $\mp=-\operatorname{sgn}(\sigma\scris+|\lambda|\scrim)$. In the case $\sigma\scris+|\lambda|\scrim>0$, this reproduces the result presented in \cite{Campoleoni:2015qrh} upon identification of inverse temperature\footnote{Note that the temperature here is rescaled by $\sigma^2$ since $|\lambda|$, being the radius of the cosmological horizon, is related to the Hawking temperature by $T_\text{H}=\sigma^2/(2\pi|\lambda|)$.} $\beta=2\pi|\lambda|$ and angular potential $\theta=4\pi\i\sigma$, as well as a rescaling of $\scris$ by a factor of two.\footnote{This should be a consequence of working with $\mathit{SL}(2,\reals)$ on the group level, which is the double (but not universal) cover of $\mathit{SO}(2,1)$.}
\subsection{The Wilson Spool and Comments on Lower Dimensions}\label{subsec:spool_def}
Now all ingredients have been provided to define the Wilson-spool operator $\mathds{W}$. This can be achieved by expressing the completely regularised partition function \eqref{eq:part_fct_reg_complete} in terms of the representation character.
\minisec{Definition of the Wilson Spool}
Gauge invariance of the trace in \eqref{eq:lnZ} can be used to express the group element in terms of the gauge fields $\omega$ and $e$, making apparent that it is basically the holonomy of the Chern-Simons field with a weight of $\i\alpha/(2\pi)$. The last non-trivial step is to claim the expression to be valid even if the gauge fields are not constant -- recall that we were exclusively dealing with flat-space cosmologies, above --, which is indicated by replacing the exponential by the path-ordered exponential. Then, writing the renormalised partition function as $Z=\exp(\mathds{W})$ we identify
\begin{align}\label{eq:def_Wilson_spool}
    \mathds{W}[\omega,e]=\frac{1}{2}\int\limits_{\mathscr{C}_\pm}\frac{\d\alpha}{\alpha}\coth\left(\frac{\alpha}{2}\right)\tr\left[\mathcal{P}\!\exp\left(\frac{\i\alpha}{2\pi}\oint(\omega+e)\right)\right]\,.
\end{align}
The trace is to be taken in an induced representation of mass $\scrim$ and spin $\scris$ of the Poincaré group. As stated previously, the integration contour $\mathscr{C}_\pm$ runs from left to right slightly above or below the real axis, that is from $-\infty\pm\i\varepsilon$ to $\infty\pm\i\varepsilon$, depending on the fall-off properties of the exponential as $|\alpha|\rightarrow\infty$. Eventually, the integral collects only the residues (stemming from the weight function) along either part of the imaginary $\alpha$-axis.

The way the Wilson spool is written suggests that it is itself a purely algebraic object that can be defined in any gauge theory, in particular we see that it is of the same form for any sign of the cosmological constant in three-dimensional gravity. Accordingly, one can make sense of expression \eqref{eq:def_Wilson_spool} in terms of a vanishing-cosmological constant limit of its (A)dS counterparts, as well. The limit takes place at the level of the algebra that the gauge fields take values in and its representation,\footnote{It is for example well known that UIRs of $\iso(2,1)$ can be obtained as a so-called ultra-relativistic (or \emph{Carrollian}) limit from highest-weight representations of $\so(2,2)$ \cite{Campoleoni:2016vsh}.} such that the form of the Wilson spool remains unaltered while the definition of its constituents changes. It is not clear, however, whether or not there is an immediate limiting procedure in relation to the integration contour but it can be argued that it is defined implicitly: once a gauge field and a UIR are fixed, the precise form of the integration contour is implicitly given through the position of the poles in the complex $\alpha$-plane as explained above.
\minisec{Comments on Radial Reduction}
In the case of vacuum spacetimes it is clear how spaces of constant curvature can be defined as hypersurfaces embedded in Minkowski spacetime. From the opposite perspective, a $d$-dimensional Minkowski spacetime can be foliated into $(d-1)$-dimensional hyperbolic (i.e. Euclidean AdS) and de-Sitter spacetimes, which can be seen as families of hypersurfaces placed inside or outside the light cone, respectively \cite{deBoer:2003vf}. This allows generic field theories in Minkowski spacetime to be radially reduced to their negative- or positive-cosmological constant constituents, which has been coined \emph{holographic reduction} and plays an important role for example in celestial holography.

The foliation of Minkowski spacetime can be nicely realised on the level of Chern-Simons gauge fields. Taking the gauge fields \eqref{eqs:spin-connection_vielbein} of the flat-space cosmology and switching to foliation coordinates $(t,\rho,\varphi)$ through the replacements $u=t\e{-\rho}$, $r=t\sinh(\rho)$ it is possible to find a gauge transformation that `gauges out' the $t$-direction, which at the same time removes the $\rho$-direction. Being a pure translation, the transformation does only affect the vielbein and one will find that $e_\varphi\sim N$, showing that the elimination of the radial direction implies a restriction of the gauge field to $\so(2,1)$ through complete elimination of $e$ only in the case $N=0$, i.e. in the absence of a cosmological horizon.

From a group-theoretic perspective, the radial reduction is a consequence of $\so(2,1)\subset\iso(2,1)$ being a subalgebra. The latter implies that any UIR of the Poincaré algebra may be restricted to its Lorentz subalgebra, thereby rendering the representation reducible and it will decompose into a (continuous) sum of irreducible representations of $\so(2,1)$. This decomposition induces a decomposition of the trace and, accordingly, if the gauge field takes values in that subalgebra we expect the Wilson spool to decompose in the same way, as well:
\begin{align}
    \mathds{W}=\int\!\!\d h\ \mu(h)\mathds{W}_{\!h}\left[A^{\text{AdS}}\right]\,.
\end{align}
This expression is supposed to be only schematic, since different UIRs of $\so(2,1)\simeq\sl(2,\reals)$ can be involved. Here, $h$ is supposed to denote the respective label (e.g. the highest weight in the case of a highest-weight representation) and $\mu(h)$ the multiplicity.

More speculative ideas on the connection between the three- and two-dimensional theories will be given in the discussion section.
\section{Discussion}\label{sec:discussion}
This work performs the construction of the Wilson spool in the case of three-dimensional gravity with vanishing cosmological constant. Building on the behaviour of massive, spinning fields under parallel transport in the presence of a non-trivial background geometry it could be shown that similar steps as have been performed in the cases of positive and negative cosmological constant do indeed lead to an analogous expression for the Wilson spool. Along the way, a known result for the one-loop partition function of a massive, spinning field in the fixed background of a flat-space cosmology could be reproduced. Let me collect a few points of critical discussion of the steps involved as well as of the final result, before commenting on further tasks.

First, the presentation within this paper takes as a starting point the transformation behaviour of massive fields in terms of the twisted-adjoint action of the Poincaré group. Although this reasoning may appear somewhat unusual, it is the impression of the author that the role of that action, which is usually not pronounced in the literature, indeed leads to a clarification of the whole minimal-coupling procedure. It should however be pointed out that the presence of the twist is not at all relevant for the calculations of the Wilson-spool construction in section \ref{sec:Wilson_spool} since only one of the two Wilson-line operators (with or without twist action) enters the construction; which one, is inconsequential since the respective sign change can be completely absorbed into a re-definition of the translation generators.

A curious observation that has been made along the derivation of the integration contour of the Wilson spool concerned the fall-off behaviour of the integrand in dependence on a parameter that combines mass aspect $M$ and angular-momentum aspect $N$ of the cosmology with mass $\scrim$ and spin $\scris$ of the field. Namely there is a range of parameters in which neither choice of how to close the integration contour is enforced but, due to a sign difference in the residues, the two possibilities will lead to slightly different results for the partition function. The standpoint of the main text is that this can be seen as an intermediate problem due to not having entirely fixed the right contour, yet. However, one may wonder if there could be any physical significance to that particular combination of parameters.

Let me now comment on a couple of possible next steps in connection to this work. First, in the case of the (A)dS Wilson spool a connection could be drawn to the Selberg zeta-function method to calculate partition functions \cite{Haupfear:2025grb} and it should be instructive to clarify whether or not the flat-space spool is similarly related to the \emph{generalised} Selberg zeta-function approach presented in \cite{Bagchi:2023ilg}. Clearly, a relation to heat-kernel methods \cite{Barnich:2015mui,Campoleoni:2015qrh} or to the recently presented calculations based on quasi-normal modes \cite{Bagchi:2025erx} would be interesting, as well. 

It has been beyond the scope of this paper to attempt a calculation of quantum-gravitational corrections to the one-loop partition function, which in the language of the Wilson spool amounts to finding contributions of higher order in perturbation theory around small gravitational constant or even non-perturbative contributions to the Chern-Simons path integral in $\langle\mathds{W}\rangle_\text{grav}$. Starting point here is the claim that the expression for the Wilson spool \eqref{eq:def_Wilson_spool} is appropriate for solutions away from the classical saddle of flat-space cosmologies as well as for geometries that are off-shell. Unfortunately, the non-compact nature of the gauge group could make an actual evaluation of the path integral difficult.

Treating anyonic fields in three spacetime dimensions does not appear much harder than the case of scalar fields, which is not surprising knowing that the partition function can be written as a combination of Poincaré characters and those simply contain the character of $\so(2)$ as a factor. A generalisation of the construction to tensor or spinor fields should however complicate the situation and it may be interesting to see how these cases work out.

Another interesting attempt will be the generalisation away from classical Einstein gravity towards supergravity or higher-spin backgrounds, which would allow to see how the behaviour of massive fields is modified in the presence of the respective charges. In both cases Chern-Simons formulations do exist \cite{Barnich:2014cwa,Afshar:2013vka}, albeit in the higher-spin case there are still several technical problems to overcome: although the construction of a particular higher-spin symmetry algebra that is suited to describe the propagation of massive degrees of freedom, along with a proposed definition for asymptotic flatness, has been introduced in the literature \cite{Ammon:2020fxs,Ammon:2022vjr,Pannier:2023srn}, it is not entirely clear how to perform finite gauge transformations due to issues with the exponentiation of Lie-algebra elements.

Within the present work only a brief comment could be given about radial reduction at the end of section \ref{sec:Wilson_spool} and this regarded the purely representation-theoretic statement on the decomposition of the Wilson spool when its input gauge field is restricted to the Lorentz subalgebra. From a geometric standpoint, a holographic foliation of spacetime (in the spirit of \cite{deBoer:2003vf}) obviously seizes to be possible in situations where the metric is Minkowski only asymptotically (such as black holes or flat-space cosmologies), but one may wonder whether a connection can be upheld within the gauge-theoretic formulation, at least in three-dimensional gravity. This hope stems from the observations that (a) the vielbein formulation utilises a basis of (co-)tangent space that is Minkowski, (b) flat-space cosmologies are quotients of Minkowski spacetime and the gauge theory is still a gauge theory of the Poincaré group, possessing hyperbolic and dS isometries as subgroups, and (c) we may take advantage of large gauge transformations. Since the gauge fields are locally pure gauge, it is clear that the vielbein $e$, which captures all contributions of translations, can be removed completely through some gauge transformation, but this does not immediately imply any statement about topological quantities such as the holonomy operator or the Wilson spool. For more rigorous and explicit statements it will be valuable to further develop suitable realisations of massive, spinning UIRs of the Poincaré algebra.

From a general perspective, there are still a number of open tasks in three-dimensional flat-space holography and in connection with massive fields as probes of (quantum) spacetime: a clear understanding of the point-particle limit of the Wilson spool, its connection to Wilson lines and Wilson-line networks as topological probes as well as a treatment of back-reaction of probes on the geometry, and the incorporation of large-central-charge corrections are just a few examples. With an eye on both Carrollian and celestial holography of flat spacetimes it will certainly prove useful to further advance these matters.
\section*{Acknowledgements}
I would like to thank Martin Ammon, Charlotte Sleight, and Massimo Taronna for general discussions, as well as Xavier Bekaert for drawing my attention towards the twisted-adjoint action.

Moreover, I would like to express my gratitude to the TPI Jena for its continuing hospitality and in particular the nice people of office 310b -- Jakob Hollweck, Christoph Sieling, and Katharina Wölfl --, with whom I enjoyed many insightful conversations.

This research was supported by the European Union\footnote{Views and opinions expressed are however those of the author only and do not necessarily reflect those of the European Union or the European Research Council. Neither the European Union nor the granting authority can be held responsible for them.} (ERC grant ``HoloBoot'', project number 101125112), by the MUR-PRIN grant No.\,PRIN2022BP52A (European Union -- Next Generation EU) and by the INFN initiative STEFI.
%
%
%
%
%
%
%
%
%
\newpage
\appendix
\addtocontents{toc}{\protect\setcounter{tocdepth}{1}}
\section{More on Wilson Lines}\label{app:sec:Wilson_lines}
Presuming that not every reader may be well-acquainted with the subject, this appendix is supposed to collect some necessary information on Wilson lines relevant for the main text and spells out conventions and notations in detail.
\minisec{Generalities}
For the moment, let us ignore possible non-trivial topological properties of the geometry on which the Wilson line is placed, i.e. assume a manifold that is simply connected. Problems appearing in the non-simply-connected case will be commented on in the end.

First, the transformation behaviour of gauge and matter fields has to be fixed; gauge invariance of the equation of motion $\d A+A\wedge A=0$ requires $A\mapsto gAg^{-1}+g\d g^{-1}$ and, in order to obtain a well-defined covariant derivative in which the gauge field contributes with the twisted-adjoint action from the left, i.e. $\operatorname{D}=\d\,+\tensor*[^{\tau}]{\ad}{_{A}}$\,, matter fields need to obey $\phi\mapsto g\phi\tau(g^{-1})$.

Now, conventions for the Wilson line can be set. First, the notation of arguments in $W(x,x_0)$ is ordered, such that a field at $x_0$ on the right-hand side is transported to $x$ on the left-hand side. Accordingly, the transformation behaviour needs to be $W(x,x_0)\mapsto g(x)W(x,x_0)g(x_0)^{-1}$. Secondly, the expansions of the path-ordered and the anti-path-ordered exponential are conventionally written as
\begin{subequations}\label{eqs:path_ordering_conv}
\begin{align}
    \Phi[B(x)]&=\left(\mathcal{P}\!\exp\int\limits_{x_0}^x\!B\right)=\mathds{1}+\int\limits_{x_0}^x\!B(x')+\int\limits_{x_0}^x\int\limits_{x_0}^{x'}\!B(x')B(x'')+\dots\,,\label{eq:path_ordering_conv}\\
    \overline{\Phi}[B(x)]&=\left(\overline{\mathcal{P}}\!\exp\int\limits_{x_0}^x\!B\right)=\mathds{1}+\int\limits_{x_0}^x\!B(x')+\int\limits_{x_0}^x\int\limits_{x_0}^{x'}\!B(x'')B(x')+\dots\,.
\end{align}
\end{subequations}
The defining differential equations of these two functionals are respectively
\begin{align}
    \d \Phi[B(x)]\,\Phi[B(x)]^{-1}=B(x)\,, && \text{and} && \overline{\Phi}[B(x)]^{-1}\d\overline{\Phi}[B(x)]=B(x)\,,
\end{align}
both augmented with the initial-value condition $\Phi[B(x=x_0)]=\mathds{1}=\overline{\Phi}[B(x=x_0)]$.

Returning to the attempt of giving an explicit form to the Wilson line, we have two possibilities to write a Cartan-Maurer element and demand it to be an element $B(x)$ of the isometry Lie algebra, leading to two possible ways to express the solution:
\begin{subequations}
\begin{align}
    \d W(x,x_0)\,W(x,x_0)^{-1}&=B(x) && \Rightarrow && W(x,x_0)=\Phi[B(x)]\,,\\
    W(x_0,x)^{-1} \d W(x_0,x)&=B(x)  && \Rightarrow && W(x_0,x)=\overline{\Phi}[B(x)]\,,
\end{align}
\end{subequations}
where the differential acts on the variable $x$ in both cases. Gauge transformation of the differential equations in both cases leads to the behaviour of $B(x)$ and thus to its identification with the background gauge field as $B(x)=-A(x)$ in the first (path-ordered) case as well as $B(x)=A(x)$ in the second (anti-path-ordered) case. An immediate implication of these statements is the simple and physically reasonable identity $W(x,x_0)=W(x_0,x)^{-1}$. To summarise, the expression for the Wilson line reads
\begin{align}
    W(x,x_0)=\mathcal{P}\!\exp\int\limits_{x_0}^x\!(-A)=\overline{\mathcal{P}}\!\exp\int\limits_{x}^{x_0}\!A\,.
\end{align}
As another remark, note that it follows from the definition given above that Wilson lines can be glued together, in the sense that $W(x,x')W(x',x_0)=W(x,x_0)$.

Finally, it should be checked that the convention for the covariant derivative already introduced above actually follows from its definition in terms of parallel transport, the latter being
\begin{align}\label{eq:def_cov_deriv}
        \operatorname{D}_\mu\phi(x):=\lim\limits_{\delta\!x\rightarrow 0}\left(\frac{\phi(x+\delta\!x)-\operatorname{T}_{x\rightarrow x+\delta\!x}\left[\phi(x)\right]}{\delta\!x^\mu}\right)\,.
\end{align}
Indeed, recalling the operator of parallel transport being defined as $\operatorname{T}_{x_0\rightarrow x}[\phi(x_0)]=W(x,x_0)\cdot\phi(x_0)\cdot\tau(W(x_0,x))$, one finds $\operatorname{D}\!\phi=\d\phi+\!\tensor*[^{\tau}]{\ad}{_{\!A}}(\phi)$, as it should be.
\minisec{Special Cases}
Clearly, in the special case when $A$ is constant, the path-ordered exponential reduces to a standard exponential; similarly, path ordering becomes obsolete in the case that $A$ takes values in an abelian algebra.

Furthermore, it is of interest to study the case of an algebra that contains an abelian ideal, such as Poincaré: We may then split off the abelian part by decomposing into spin connection and vielbein, $A=\omega+e$. Then the Wilson line may be decomposed as $W(x,x_0)=W_{\text{L}}(x,x_0)\cdot W_{\text{T}}(x,x_0)$, the two factors taking values in the subgroups of Lorentz transformations and translations, respectively. Only the Lorentz part will require path ordering while the abelian part is an ordinary exponential,
\begin{align}
    W_{\text{L}}(x,x_0)=\mathcal{P}\!\exp\int\limits_{x_0}^x(-\omega)\,, && W_{\text{T}}(x,x_0)=\e{-\xi_{\text{T}}(x,x_0)}\,,
\end{align}
where the Lie-algebra element can explicitly be given as follows:
\begin{align}
    \xi_{\text{T}}(x,x_0)=\int\limits_{x_0}^x W_{\text{L}}(x_0,x')\cdot e(x')\cdot W_{\text{L}}(x',x_0)\,.
\end{align}
An advantage of this decomposition is that the action of the twist automorphism on purely translational group elements can be immediately identified with the inverse. The behaviour of both factors under gauge transformations $g(x)=g_{\text{L}}(x)g_{\text{T}}(x)$, where $g_{\text{T}}=\exp(\xi_{\text{T}})$, is found to be
\begin{subequations}
\begin{align}
    W_{\text{L}}(x,x_0)&\mapsto g_{\text{L}}(x)W_{\text{L}}(x,x_0)g_{\text{L}}(x_0)^{-1}\,,\\
    W_{\text{T}}(x,x_0)&\mapsto g_{\text{L}}(x_0)\left(\mathds{1}-\xi_{\text{T}}(x)+\operatorname{T}_{x\rightarrow x_0}\left[\xi_{\text{T}}(x)\right]\right)g_{\text{T}}(x)W_{\text{T}}(x,x_0)g_{\text{T}}(x_0)^{-1}g_{\text{L}}(x_0)^{-1}\,.
\end{align}
\end{subequations}
As expected, the Lorentzian part neatly reduces to an element of the $\mathit{SO}(2,1)$-subgroup; the translational part takes on a more involved form but transforms homogeneously under the adjoint action of Lorentz transformations, localised at $x_0$, and therein the parallel-transport operator $\operatorname{T}_{x\rightarrow x_0}\left[\xi_{\text{T}}(x)\right]=W_{\text{L}}(x_0,x)\cdot\xi_{\text{T}}(x)\cdot W_{\text{L}}(x,x_0)$ is to be understood in terms of the original, untransformed Wilson line. A decomposition like that leads to an alternative form of the Wilson-spool operator in flat space, perhaps being of use in future applications.

Finally, properties of the flat connection $A$ may be exploited, namely its being pure gauge. Any solution of $\d A+A\wedge A=0$ can be written as $A=g_0^{-1}\d g_0$ for a specific group element $g_0$ (which, apparently, is given in terms of a path-ordered exponential of $A$), implying that it lies on the same gauge orbit as the trivial solution $A=0$. Then from the transformation behaviour of the Wilson line follows that it can be expressed solely in terms of the (large) gauge transformation that connects a given $A$ to the trivial solution, namely $W(x,x_0)=g_0(x)^{-1}g_0(x_0)$. However, keep in mind that we are still assuming a simply connected manifold here, such that the closed Wilson line, $x\rightarrow x_0$, always equals the unit element, i.e. the geometry has a trivial holonomy.
\minisec{Non-Trivial Holonomy}
In the case of a manifold that is not simply connected, the definition of the path-ordered exponential in terms of a differential equation given above is no longer complete, since paths connecting the same end points cannot be continuously transformed into each other any more. In that case one may formally work with the same gauge-transformation behaviour of the Wilson line as before with the additional condition of $g=g(x)$ not being a single-valued function but rather carrying an additional representation $b$ of the fundamental group $\pi_1(\mathcal{M})$ of the manifold $\mathcal{M}$, i.e. $g(x)\simeq g(x)\circ b$. In the case of $\pi_1(\mathcal{M})=\mathds{Z}$, such as for the BTZ black hole or flat-space cosmologies, such a representation looks like $k\mapsto b(k)\in G$ with $G$ being the isometry group and $k\in\mathds{Z}$ denoting a winding number. Considering closed Wilson lines, the respective group element of the representative $k=1$ is associated with the non-trivial holonomy around the non-contractible cycle. This presents an obstacle in the attempt to connect the three-dimensional Wilson spool at vanishing cosmological constant with its two-dimensional (A)dS counterparts through a large gauge transformation, as discussed in the main text.
\printbibliography

@article{Afshar:2013bla,
    author = "Afshar, Hamid R.",
    title = "{Flat/AdS boundary conditions in three dimensional conformal gravity}",
    eprint = "1307.4855",
    archivePrefix = "arXiv",
    primaryClass = "hep-th",
    doi = "10.1007/JHEP10(2013)027",
    journal = "JHEP",
    volume = "10",
    pages = "027",
    year = "2013"
}

@article{Prokushkin:1998bq,
    author = "Prokushkin, S. F. and Vasiliev, Mikhail A.",
    title = "{Higher spin gauge interactions for massive matter fields in 3-D AdS space-time}",
    eprint = "hep-th/9806236",
    archivePrefix = "arXiv",
    reportNumber = "FIAN-TD-16-98",
    doi = "10.1016/S0550-3213(98)00839-6",
    journal = "Nucl. Phys. B",
    volume = "545",
    pages = "385",
    year = "1999"
}

@article{Kessel:2018zqm,
    author = "Kessel, Pan and Raeymaekers, Joris",
    title = "{Simple unfolded equations for massive higher spins in AdS$_{3}$}",
    eprint = "1805.07279",
    archivePrefix = "arXiv",
    primaryClass = "hep-th",
    doi = "10.1007/JHEP08(2018)076",
    journal = "JHEP",
    volume = "08",
    pages = "076",
    year = "2018"
}

@article{Ammon:2020fxs,
    author = "Ammon, Martin and Pannier, Michel and Riegler, Max",
    title = "{Scalar Fields in 3D Asymptotically Flat Higher-Spin Gravity}",
    eprint = "2009.14210",
    archivePrefix = "arXiv",
    primaryClass = "hep-th",
    doi = "10.1088/1751-8121/abdbc6",
    journal = "J. Phys. A",
    volume = "54",
    number = "10",
    pages = "105401",
    year = "2021"
}

@article{Campoleoni:2016vsh,
    author = "Campoleoni, Andrea and Gonzalez, Hernan A. and Oblak, Blagoje and Riegler, Max",
    editor = "Brink, Lars and Henneaux, Marc and Vasiliev, Mikhail A.",
    title = "{BMS Modules in Three Dimensions}",
    eprint = "1603.03812",
    archivePrefix = "arXiv",
    primaryClass = "hep-th",
    doi = "10.1142/S0217751X16500688",
    journal = "Int. J. Mod. Phys. A",
    volume = "31",
    number = "12",
    pages = "1650068",
    year = "2016"
}

@article{Campoleoni:2015qrh,
    author = "Campoleoni, Andrea and Gonzalez, Hernan A. and Oblak, Blagoje and Riegler, Max",
    title = "{Rotating Higher Spin Partition Functions and Extended BMS Symmetries}",
    eprint = "1512.03353",
    archivePrefix = "arXiv",
    primaryClass = "hep-th",
    doi = "10.1007/JHEP04(2016)034",
    journal = "JHEP",
    volume = "04",
    pages = "034",
    year = "2016"
}

@article{Afshar:2013vka,
      author         = "Afshar, Hamid and Bagchi, Arjun and Fareghbal, Reza and Grumiller, Daniel and Rosseel, Jan",
      title          = "{Spin-3 Gravity in Three-Dimensional Flat Space}",
      journal        = "Phys. Rev. Lett.",
      volume         = "111",
      year           = "2013",
      number         = "12",
      pages          = "121603",
      doi            = "10.1103/PhysRevLett.111.121603",
      eprint         = "1307.4768",
      archivePrefix  = "arXiv",
      primaryClass   = "hep-th",
      reportNumber   = "TUW-13-09",
      SLACcitation   = "%%CITATION = ARXIV:1307.4768;%%"
}

@article{Ammon:2011ua,
      author         = "Ammon, Martin and Kraus, Per and Perlmutter, Eric",
      title          = "{Scalar fields and three-point functions in $D=3$ higher
                        spin gravity}",
      journal        = "JHEP",
      volume         = "07",
      year           = "2012",
      pages          = "113",
      doi            = "10.1007/JHEP07(2012)113",
      eprint         = "1111.3926",
      archivePrefix  = "arXiv",
      primaryClass   = "hep-th",
      SLACcitation   = "%%CITATION = ARXIV:1111.3926;%%"
}

@article{Vasiliev:1999ba,
      author         = "Vasiliev, Mikhail A.",
      title          = "{Higher spin gauge theories: Star product and AdS space}",
      year           = "1999",
      pages          = "533-610",
      doi            = "10.1142/9789812793850_0030",
      eprint         = "hep-th/9910096",
      archivePrefix  = "arXiv",
      primaryClass   = "hep-th",
      reportNumber   = "FIAN-TD-24-99",
      SLACcitation   = "%%CITATION = HEP-TH/9910096;%%"
}

@article{Achucarro:1986uwr,
    author = "Achucarro, A. and Townsend, P. K.",
    editor = "Salam, A. and Sezgin, E.",
    title = "{A Chern-Simons Action for Three-Dimensional anti-De Sitter Supergravity Theories}",
    reportNumber = "Print-87-0078 (CAMBRIDGE)",
    doi = "10.1016/0370-2693(86)90140-1",
    journal = "Phys. Lett. B",
    volume = "180",
    pages = "89",
    year = "1986"
}

@article{Witten:1988hc,
    author = "Witten, Edward",
    title = "{(2+1)-Dimensional Gravity as an Exactly Soluble System}",
    reportNumber = "IASSNS-HEP-88-32",
    doi = "10.1016/0550-3213(88)90143-5",
    journal = "Nucl. Phys. B",
    volume = "311",
    pages = "46",
    year = "1988"
}

@article{Witten:2007kt,
    author = "Witten, Edward",
    title = "{Three-Dimensional Gravity Revisited}",
    eprint = "0706.3359",
    archivePrefix = "arXiv",
    primaryClass = "hep-th",
    month = "6",
    year = "2007"
}

@article{Barnich:2012aw,
    author = "Barnich, Glenn and Gomberoff, Andres and Gonzalez, Hernan A.",
    title = "{The Flat limit of three dimensional asymptotically anti-de Sitter spacetimes}",
    eprint = "1204.3288",
    archivePrefix = "arXiv",
    primaryClass = "gr-qc",
    doi = "10.1103/PhysRevD.86.024020",
    journal = "Phys. Rev. D",
    volume = "86",
    pages = "024020",
    year = "2012"
}

@phdthesis{Oblak:2016eij,
    author = "Oblak, Blagoje",
    title = "{BMS Particles in Three Dimensions}",
    eprint = "1610.08526",
    archivePrefix = "arXiv",
    primaryClass = "hep-th",
    doi = "10.1007/978-3-319-61878-4",
    school = "Brussels U.",
    year = "2016"
}

@article{Ammon:2022vjr,
    author = "Ammon, Martin and Pannier, Michel",
    title = "{Unfolded Fierz-Pauli equations in three-dimensional asymptotically flat spacetimes}",
    eprint = "2211.12530",
    archivePrefix = "arXiv",
    primaryClass = "hep-th",
    doi = "10.1007/JHEP02(2023)161",
    journal = "JHEP",
    volume = "02",
    pages = "161",
    year = "2023"
}

@phdthesis{Pannier:2023srn,
    author = "Pannier, Michel",
    title = "{Aspects of holography and higher spins in three-dimensional asymptotically flat spacetimes}",
    school = "U. Jena (main)",
    url = 	{https://www.db-thueringen.de/receive/dbt_mods_00056722},
    year = "2023"
}

@article{Barnich:2014cwa,
    author = "Barnich, Glenn and Donnay, Laura and Matulich, Javier and Troncoso, Ricardo",
    title = "{Asymptotic symmetries and dynamics of three-dimensional flat supergravity}",
    eprint = "1407.4275",
    archivePrefix = "arXiv",
    primaryClass = "hep-th",
    reportNumber = "CECS-PHY-14-02",
    doi = "10.1007/JHEP08(2014)071",
    journal = "JHEP",
    volume = "08",
    pages = "071",
    year = "2014"
}

@article{Cornalba:2002fi,
    author = "Cornalba, Lorenzo and Costa, Miguel S.",
    title = "{A New cosmological scenario in string theory}",
    eprint = "hep-th/0203031",
    archivePrefix = "arXiv",
    reportNumber = "ITFA-2002-07, LPTENS-02-19",
    doi = "10.1103/PhysRevD.66.066001",
    journal = "Phys. Rev. D",
    volume = "66",
    pages = "066001",
    year = "2002"
}

@article{Alkalaev:2019xuv,
    author = "Alkalaev, Konstantin and Bekaert, Xavier",
    title = "{Towards higher-spin AdS$_2$/CFT$_1$ holography}",
    eprint = "1911.13212",
    archivePrefix = "arXiv",
    primaryClass = "hep-th",
    doi = "10.1007/JHEP04(2020)206",
    journal = "JHEP",
    volume = "04",
    pages = "206",
    year = "2020"
}

@article{Alkalaev:2020kut,
    author = "Alkalaev, Konstantin and Bekaert, Xavier",
    title = "{On BF-type higher-spin actions in two dimensions}",
    eprint = "2002.02387",
    archivePrefix = "arXiv",
    primaryClass = "hep-th",
    doi = "10.1007/JHEP05(2020)158",
    journal = "JHEP",
    volume = "05",
    pages = "158",
    year = "2020"
}

@article{Castro:2023dxp,
    author = "Castro, Alejandra and Coman, Ioana and Fliss, Jackson R. and Zukowski, Claire",
    title = "{Keeping matter in the loop in dS$_{3}$ quantum gravity}",
    eprint = "2302.12281",
    archivePrefix = "arXiv",
    primaryClass = "hep-th",
    doi = "10.1007/JHEP07(2023)120",
    journal = "JHEP",
    volume = "07",
    pages = "120",
    year = "2023",
    note = "[Erratum: JHEP 09, 004 (2024)]"
}

@article{Castro:2023bvo,
    author = "Castro, Alejandra and Coman, Ioana and Fliss, Jackson R. and Zukowski, Claire",
    title = "{Coupling Fields to 3D Quantum Gravity via Chern-Simons Theory}",
    eprint = "2304.02668",
    archivePrefix = "arXiv",
    primaryClass = "hep-th",
    doi = "10.1103/PhysRevLett.131.171602",
    journal = "Phys. Rev. Lett.",
    volume = "131",
    number = "17",
    pages = "171602",
    year = "2023"
}

@article{Bagchi:2013lma,
    author = "Bagchi, Arjun and Detournay, Stephane and Grumiller, Daniel and Simon, Joan",
    title = "{Cosmic Evolution from Phase Transition of Three-Dimensional Flat Space}",
    eprint = "1305.2919",
    archivePrefix = "arXiv",
    primaryClass = "hep-th",
    reportNumber = "TUW-13-06",
    doi = "10.1103/PhysRevLett.111.181301",
    journal = "Phys. Rev. Lett.",
    volume = "111",
    number = "18",
    pages = "181301",
    year = "2013"
}

@article{Bagchi:2025erx,
    author = "Bagchi, Arjun and Biswas, Supratik and Kakkar, Astha and Mondal., Saikat",
    title = "{Scalar fields and 3D Flat Space Cosmologies}",
    eprint = "2506.02148",
    archivePrefix = "arXiv",
    primaryClass = "hep-th",
    month = "6",
    year = "2025"
}

@article{Fliss:2025sir,
    author = "Fliss, Jackson R.",
    title = "{Massive fields and Wilson spools in JT gravity}",
    eprint = "2503.08657",
    archivePrefix = "arXiv",
    primaryClass = "hep-th",
    month = "3",
    year = "2025"
}

@article{Denef:2009kn,
    author = "Denef, Frederik and Hartnoll, Sean A. and Sachdev, Subir",
    title = "{Black hole determinants and quasinormal modes}",
    eprint = "0908.2657",
    archivePrefix = "arXiv",
    primaryClass = "hep-th",
    reportNumber = "NSF-KITP-09-145",
    doi = "10.1088/0264-9381/27/12/125001",
    journal = "Class. Quant. Grav.",
    volume = "27",
    pages = "125001",
    year = "2010"
}

@article{Kitaev:2017hnr,
    author = "Kitaev, Alexei",
    title = "{Notes on $\widetilde{\mathrm{SL}}(2,\mathbb{R})$ representations}",
    eprint = "1711.08169",
    archivePrefix = "arXiv",
    primaryClass = "hep-th",
    month = "11",
    year = "2017"
}

@article{Bagchi:2013qva,
    author = "Bagchi, Arjun and Basu, Rudranil",
    title = "{3D Flat Holography: Entropy and Logarithmic Corrections}",
    eprint = "1312.5748",
    archivePrefix = "arXiv",
    primaryClass = "hep-th",
    doi = "10.1007/JHEP03(2014)020",
    journal = "JHEP",
    volume = "03",
    pages = "020",
    year = "2014"
}

@article{deBoer:2003vf,
    author = "de Boer, Jan and Solodukhin, Sergey N.",
    title = "{A Holographic reduction of Minkowski space-time}",
    eprint = "hep-th/0303006",
    archivePrefix = "arXiv",
    reportNumber = "ITFA-2003-11",
    doi = "10.1016/S0550-3213(03)00494-2",
    journal = "Nucl. Phys. B",
    volume = "665",
    pages = "545--593",
    year = "2003"
}

@book{Carlip:1998uc,
    author = "Carlip, Steven",
    title = "{Quantum gravity in 2+1 dimensions}",
    doi = "10.1017/CBO9780511564192",
    isbn = "978-0-521-54588-4",
    publisher = "Cambridge University Press",
    series = "Cambridge Monographs on Mathematical Physics",
    month = "12",
    year = "2003"
}

@article{Dzhordzhadze:1994np,
    author = "Dzhordzhadze, G. and O'Raifeartaigh, L. and Tsutsui, I.",
    title = "{Quantization of a relativistic particle on the SL(2,R) manifold based on Hamiltonian reduction}",
    eprint = "hep-th/9407059",
    archivePrefix = "arXiv",
    reportNumber = "INS-1042, DIAS-STP-94-22",
    doi = "10.1016/0370-2693(94)90549-5",
    journal = "Phys. Lett. B",
    volume = "336",
    pages = "388--394",
    year = "1994"
}

@article{Ammon:2013hba,
    author = "Ammon, Martin and Castro, Alejandra and Iqbal, Nabil",
    title = "{Wilson Lines and Entanglement Entropy in Higher Spin Gravity}",
    eprint = "1306.4338",
    archivePrefix = "arXiv",
    primaryClass = "hep-th",
    reportNumber = "NSF-KITP-13-112",
    doi = "10.1007/JHEP10(2013)110",
    journal = "JHEP",
    volume = "10",
    pages = "110",
    year = "2013"
}

@article{Castro:2018srf,
    author = "Castro, Alejandra and Iqbal, Nabil and Llabr{\'e}s, Eva",
    title = "{Wilson lines and Ishibashi states in AdS$_{3}$/CFT$_{2}$}",
    eprint = "1805.05398",
    archivePrefix = "arXiv",
    primaryClass = "hep-th",
    doi = "10.1007/JHEP09(2018)066",
    journal = "JHEP",
    volume = "09",
    pages = "066",
    year = "2018"
}

@article{Susskind:1998vk,
    author = "Susskind, Leonard",
    editor = "Burgess, C. P. and Myers, Robert C.",
    title = "{Holography in the flat space limit}",
    eprint = "hep-th/9901079",
    archivePrefix = "arXiv",
    doi = "10.1063/1.1301570",
    journal = "AIP Conf. Proc.",
    volume = "493",
    number = "1",
    pages = "98--112",
    year = "1999"
}

@inproceedings{levy1965nouvelle,
  title={Une nouvelle limite non-relativiste du groupe de Poincar{\'e}},
  author={L{\'e}vy-Leblond, Jean-Marc},
  booktitle={Annales de l'IHP Physique th{\'e}orique},
  volume={3},
  number={1},
  pages={1--12},
  year={1965}
}

@article{sen1966analogue,
  title={On an analogue of the Galilei group},
  author={Sen Gupta, ND},
  journal={Il Nuovo Cimento A (1965-1970)},
  volume={44},
  number={2},
  pages={512--517},
  year={1966},
  publisher={Springer}
}

@article{Barnich:2010eb,
    author = "Barnich, Glenn and Troessaert, Cedric",
    title = "{Aspects of the BMS/CFT correspondence}",
    eprint = "1001.1541",
    archivePrefix = "arXiv",
    primaryClass = "hep-th",
    reportNumber = "ULB-TH-09-28",
    doi = "10.1007/JHEP05(2010)062",
    journal = "JHEP",
    volume = "05",
    pages = "062",
    year = "2010"
}

@article{Bagchi:2010zz,
    author = "Bagchi, Arjun",
    title = "{Correspondence between Asymptotically Flat Spacetimes and Nonrelativistic Conformal Field Theories}",
    eprint = "1006.3354",
    archivePrefix = "arXiv",
    primaryClass = "hep-th",
    doi = "10.1103/PhysRevLett.105.171601",
    journal = "Phys. Rev. Lett.",
    volume = "105",
    pages = "171601",
    year = "2010"
}

@article{Bourne:2024ded,
    author = "Bourne, Robert and Castro, Alejandra and Fliss, Jackson R.",
    title = "{Spinning up the spool: massive spinning fields in 3d quantum gravity}",
    eprint = "2407.09608",
    archivePrefix = "arXiv",
    primaryClass = "hep-th",
    doi = "10.1088/1751-8121/ad9e55",
    journal = "J. Phys. A",
    volume = "58",
    number = "2",
    pages = "025402",
    year = "2025"
}

@book{Cherednikov:2020mtu,
    author = "Cherednikov, Igor Olegovich and Mertens, Tom and Van der Veken, Frederik",
    title = "{Wilson Lines in Quantum Field Theory}",
    doi = "10.1515/9783110651690",
    isbn = "978-3-11-065092-1",
    publisher = "De Gruyter",
    series = "De Gruyter Studies in Mathematical Physics",
    volume = "24",
    month = "1",
    year = "2020"
}

@article{Bourne:2025azc,
    author = "Bourne, Robert and Fliss, Jackson R. and Knighton, Bob",
    title = "{A spool for every quotient: One-loop partition functions in AdS$_3$ gravity}",
    eprint = "2507.05364",
    archivePrefix = "arXiv",
    primaryClass = "hep-th",
    month = "7",
    year = "2025"
}

@article{Haupfear:2025grb,
    author = "Haupfear, Samuel and Martin, Victoria and Svesko, Andrew and Zukowski, Claire",
    title = "{Untangling Selberg from the Wilson spool: 1-loop determinants and trace formulae in (A)dS$_{3}$}",
    eprint = "2507.05358",
    archivePrefix = "arXiv",
    primaryClass = "hep-th",
    month = "7",
    year = "2025"
}

@article{Barnich:2015mui,
    author = "Barnich, Glenn and Gonzalez, Hernan A. and Maloney, Alexander and Oblak, Blagoje",
    title = "{One-loop partition function of three-dimensional flat gravity}",
    eprint = "1502.06185",
    archivePrefix = "arXiv",
    primaryClass = "hep-th",
    doi = "10.1007/JHEP04(2015)178",
    journal = "JHEP",
    volume = "04",
    pages = "178",
    year = "2015"
}

@article{Bagchi:2023ilg,
    author = "Bagchi, Arjun and Keeler, Cynthia and Martin, Victoria and Poddar, Rahul",
    title = "{A generalized Selberg zeta function for flat space cosmologies}",
    eprint = "2312.06770",
    archivePrefix = "arXiv",
    primaryClass = "hep-th",
    doi = "10.1007/JHEP04(2024)066",
    journal = "JHEP",
    volume = "04",
    pages = "066",
    year = "2024"
}

@article{Gunaydin:1986fe,
    author = "Gunaydin, M. and Sierra, G. and Townsend, P. K.",
    title = "{The Unitary Supermultiplets of $d=3$ Anti-de Sitter and $d=2$ Conformal Superalgebras}",
    reportNumber = "UCRL-93923",
    doi = "10.1016/0550-3213(86)90293-2",
    journal = "Nucl. Phys. B",
    volume = "274",
    pages = "429--447",
    year = "1986"
}
\end{document}